\documentclass[preprint]{elsarticle}
\usepackage{csquotes}
\usepackage{graphicx}
\usepackage{ulem,color}
\usepackage[dvipsnames]{xcolor/xcolor}
\usepackage{natbib}
\usepackage[bookmarks=true]{hyperref}
\usepackage{xspace}
\usepackage{amsmath}
\usepackage{url}

\usepackage[margin=1in]{geometry}

\usepackage{lineno}
\modulolinenumbers[5]
\journal{Journal of Computational Physics}
\biboptions{sort&compress}

\def\lesssim{\mathrel{\hbox{\rlap{\hbox{\lower4pt\hbox{$\sim$}}}\hbox{$<$}}}}
\def\gtrsim{\mathrel{\hbox{\rlap{\hbox{\lower4pt\hbox{$\sim$}}}\hbox{$>$}}}}

\newcommand{\bea}{\begin{eqnarray}}
\newcommand{\eea}{\end{eqnarray}}

\newcommand{\harm}{{\sc Harm3d}\xspace} 
 
\newcommand{\pw}{{\sc Patchwork}\xspace}
\newcommand{\pwmhd}{{\sc PatchworkMHD}\xspace}
\newcommand{\pwWave}{{\sc PatchworkWave}\xspace}

\def\lambdabar{%
\relax
\bgroup
\def\@tempa{\hbox{\raise.73\ht0
\hbox to0pt{\kern.25\wd0\vrule width.5\wd0
height.1pt depth.1pt\hss}\box0}}%
\mathchoice{\setbox0\hbox{$\displaystyle\lambda$}\@tempa}%
{\setbox0\hbox{$\textstyle\lambda$}\@tempa}%
{\setbox0\hbox{$\scriptstyle\lambda$}\@tempa}%
{\setbox0\hbox{$\scriptscriptstyle\lambda$}\@tempa}%
\egroup
}

\begin{document}
\begin{frontmatter}

\title{\pwWave: A Multipatch Infrastructure for Multiphysics/Multiscale/Multiframe/Multimethod\\ Simulations at Arbitrary Order}

\address[LANL1]{X Computational Physics, Los Alamos National Laboratory,
  P.O. Box 1663, Los Alamos, NM 87545}
\address[LANL2]{Center for Theoretical Astrophysics, Los Alamos National Laboratory,
  P.O. Box 1663, Los Alamos, NM 87545}
\address[RIT]{Center for Computational Relativity and Gravitation,
  Rochester Institute of Technology, Rochester, NY 14623}
\address[ORNL1]{National Center for Computational Sciences, Oak Ridge National Laboratory, 
  P.O. Box 2008, Oak Ridge, TN 37831-6164}
\address[ORNL2]{Physics Division, Oak Ridge National Laboratory, P.O. Box 2008, Oak Ridge, TN 37831-6354}
\address[TULSA]{Department of Physics and Engineering Physics,
  The University of Tulsa, Tulsa, OK 74104}
\address[NASA]{Gravitational Astrophysics Laboratory, Goddard Space Flight Center, Greenbelt, MD 20771}
\address[HS]{Harvard-Smithsonian Center for Astrophysics, Cambridge, MA 02193}
\address[RC]{Theoretical Division (T-3),
              Los Alamos National Laboratory,
              P.O. Box 1663,
              Los Alamos, NM 87545}
\address[JHU]{Department of Physics and Astronomy, Johns Hopkins University, Baltimore, MD 21218}

\author[LANL1,LANL2,RIT]{Dennis B. Bowen\corref{mycorrespondingauthor}}
\cortext[mycorrespondingauthor]{Corresponding author}
\ead{dbowen@lanl.gov}
\author[RIT]{Mark Avara}
\author[ORNL1,ORNL2,RIT]{Vassilios Mewes}
\author[RIT]{Yosef Zlochower}
\author[TULSA,NASA]{Scott C. Noble}
\author[RIT]{Manuela Campanelli}
\author[HS]{Hotaka Shiokawa}
\author[RC]{Roseanne M. Cheng}
\author[JHU]{Julian H. Krolik}

\begin{abstract}
We present an extension of the PatchworkMHD code~\citep{PWMHD},
itself an MHD-capable extension of the \pw code~\citep{PATCHWORK}, for which several algorithms
presented here were co-developed. Its purpose is to create 
a ``multipatch'' scheme compatible
with numerical simulations of arbitrary equations of motion at any
discretization order in space and time. In {the \pw framework}, the global
simulation is comprised of an arbitrary number of moving, local meshes,
or ``patches'', which are free to employ their own resolution,
coordinate system/topology, physics equations, reference frame, and in our new approach, numerical method.
Each local patch exchanges boundary data with a single global patch on which all other
patches reside through a client-router-server parallelization model.
In generalizing \pw to be compatible with arbitrary order time integration, \pwmhd and \pwWave have
significantly improved the interpatch interpolation accuracy by removing an 
interpolation of interpolated data feedback present in the original \pw code. 
Furthermore, we extend \pw to be \textit{multimethod} by allowing multiple state vectors 
to be updated simultaneously, with each
state vector providing its own interpatch interpolation and transformation
procedures. As such, our scheme is compatible with nearly any set of hyperbolic
partial differential equations. We demonstrate our changes through the implementation of
a scalar wave toy-model that is evolved on arbitrary, time dependent patch configurations
at 4th order accuracy.
\end{abstract}

\begin{keyword}
  Multipatch methods, Overset meshes, Wave-like systems
\end{keyword}

\end{frontmatter}
\section{Introduction}
\label{sec:introduction}
\subsection{Broad Applicability of Multiphysics/Multiscale Computation}
One of the greatest challenges facing modern
computational physics is the self-consistent
modeling of multiphysics/multiscale systems
at high fidelity. Particularly, because many
physical systems of interest can contain
significant heterogeneities with regard to
relevant physical processes and characteristic
timescales. Furthermore, these inhomogeneities cannot
always be cleanly partitioned into distinctly
independent regions and must often be simulated
simultaneously. Finally, because simulation cost
is generally dictated by the most expensive physical
process, shortest characteristic lengths, and longest
dynamical timescales,
over simplified models are often employed to the
detriment of the global simulation.

A simulation which 
includes multiple distinct physical processes is referred to as a ``multiphysics''
simulation. A multiphysics simulation code must therefore solve
various sets of equations  simultaneously (often requiring different sophisticated
numerical techniques), and is responsible for coupling
each physical process together. For instance, some combination of
the equations of
hydrodynamics, electromagnetism, chemical reactions, radiation transport, solid
mechanics, gravity, and diffusion processes may be coupled together
to accurately capture the physics at hand.

Multiphysics examples are ubiquitous throughout computational
physics and span from modeling terrestrial experiments to astrophysics.
For instance, simulating fusion experiments such as MagLIF~\citep{PhysRevLett.113.155003}, which requires radiation
and multi-material magnetohydrodynamics (MHD). Meanwhile, multiphysics modeling in astrophysics
includes accretion onto compact objects~\citep{2002apa..book.....F}, stellar astrophysics~\citep{2007PhR...442...38J},
and cosmology~\citep{2005Natur.435..629S}. All of which require some level of MHD,
radiation, gravity, and/or reactions at varying levels of
accuracy.

The complexity is further compounded by the variation in characteristic
length and timescales of each physical process at work. For example,
in common envelope evolution astrophysics the dynamical timescales
relevant to the immediate vicinity of the engulfed neutron star
differ from that of the global system by ten orders of magnitude~\citep{Ivanova2013}.
Additionally,
the relevant length and timescales on which a single physical process
can operate may vary significantly through the physical system of
interest. An example in engineering physics includes the modeling
of various turbulence scales near wind turbines~\citep{turbine}.
Such systems are referred to as ``multiscale'' and represent
a significant computational challenge.
Furthermore, the approximate symmetries, motions, and resolution
requirements of each physical
scale are often nonuniform through the problem domain.

In addition to the complexities described above, often the various
physics components require different numerical methods. For instance,
one application may make use of an Eulerian finite volume approach while another
makes use of finite difference, finite element, or Lagrangian methods.
We refer to such applications as ``multimethod''.

In this paper we present a new code (\pwWave) which is motivated as a proof-of-principle
first step towards a sophisticated multiphysics/multiscale/multiframe/multimethod
application in relativistic astrophysics. 
With the detection of gravitational waves by 
LIGO~\citep{GWdetect,Abbott:2016nmj,PhysRevLett.118.221101,2017ApJ...851L..35A,LIGOScientific:2018mvr,LIGOScientific:2018jsj} 
and the planned launch of LISA~\citep{2019arXiv190305287K,2019arXiv190304417B}, there exists a significant deficiency
in the knowledge of what electromagnetic counterparts to supermassive
binary black hole mergers would be. To zeroth order, the electromagnetic
counterpart will be directly related to the structure and quantity of gas in the
immediate vicinity of the black holes at merger~\citep{Krolik10}. Much effort has
recently been undertaken to ascertain the gas dynamics in SMBBHs
in both the Newtonian~\citep{MM08,Shi12,DOrazio13,Farris14,Farris15,Farris15a,Shi2015,DOrazio16,RyanMacFadyen17,2018MNRAS.476.2249T,Moody2019} 
and relativistic~\citep{2010ApJ...715.1117B,Pal10,Farris11,Bode12,Farris12,Giacomazzo12,Noble12,Gold14,Bowen17,Bowen18,Bowen19} regimes.
Through these simulations direct predictions of electromagnetic signatures
may be made~\citep{dAscoli:2018fjw}.

Recent simulations of relativistic SMBBHs seem to imply that the
electromagnetic counterpart may depend sensitively on the gravitationally
driven inspiral phase directly before merger~\citep{Bowen18,Bowen19}. Unfortunately,
there does not currently exist a simulation code capable of
evolving the merger proper on spherical grids over the long durations necessary to
capture the dynamical timescales of the larger disk
feeding material to the black holes.
(For a review of relativistic SMBBH accretion see~\cite{2019Galax...7...63G}). However, a
new code (\pwmhd), which will be described in a companion publication \citep{PWMHD},
has been developed to extend the \pw infrastructure~\citep{PATCHWORK} to include MHD
and to work with
the full \harm code (see~\cite{Noble09,Noble12,Bowen17}).
The \harm code already contains the necessary ingredients to
simulate the inspiral phase of SMBBH merger and when coupled with \pwmhd will
allow efficient modeling of this crucial phase just prior to merger.

However, \harm is not capable of modeling the SMBBH merger phase. 
Unlike most current codes which couple numerical relativity to MHD,
\harm is capable of using curvi-linear coordinates to preserve
angular momentum and scales well to large CPU counts.
Motivated by this, we wish to extend the \pwmhd infrastructure to allow full
numerical relativity coupled to MHD in the \harm code. The first step
in this goal is the creation of a multimethod capable multipatch scheme
in the \harm framework.

From here on, unless otherwise stated, we use \pw to implicitly refer to the enhanced
  version extended and implemented into the full \harm code, which will be described in \citep{PWMHD}, and on which 
  \pwmhd and \pwWave are built. 
  We refer to the version published in~\citep{PATCHWORK}
  as the ``original \pw'' and/or via citation. Finally, we reserve \pwmhd and \pwWave
  as references to application specific branches of the \pw code family. 
  
\subsection{Modeling Multiphysics/Multiscale Systems}
Many sophisticated techniques have been developed to handle the
multiscale nature of most multiphysics applications. A common approach
is the use of adaptive mesh refinement (AMR)~\citep{BERGER1984484,BERGER198964}. AMR allows
the dynamic adjustment of resolution throughout the computational domain
according to some criteria. However, this comes with several drawbacks.

First, most AMR algorithms use exact 2-to-1 mesh refinement and require
many levels of refinement to extend sufficiently large resolution
scale gaps. This often comes at the detriment to scaling to many processors.
Additionally, if left unchecked or without properly defined refinement
criteria, there is no guarantee that the AMR algorithm will detect and
sufficiently resolve the characteristic scales.
Conversely, over refinement in regions of steep
gradients, such as shocks, can increase computational cost unnecessarily.
 Finally, AMR algorithms are generally employed in a global
coordinate basis incapable of following local, physical symmetries of the system.

Greater flexibility can be employed in sophisticated AMR libraries,
such as Chombo~\citep{Adams:2015kgr}, which allow mesh refinement into
separate grids, embedded boundaries, and mapped multiblocks. The method
of splitting the computational domain into several overlapping
meshes, or ``overset meshes'', has a wealth of literature
(for a review of development at NASA see~\cite{CHAN2009496}). So-called
``Chimera'' grids~\citep{CHIMERA}, allow the deployment of physics on sophisticated
overlapping meshes~\citep{doi:10.2514/3.12078,TANG2003567,doi:10.2514/6.2014-2008,turbine,
doi:10.2514/6.2014-0907,WANG2014333,
doi:10.2514/6.2016-3571,doi:10.2514/6.2016-0797,doi:10.2514/6.2016-2963,doi:10.2514/6.2016-0814,
doi:10.2514/6.2017-0362,
doi:10.2514/6.2018-2056,doi:10.2514/6.2018-1266}. In fact, the need for overset mesh techniques
in the computational fluid dynamics community is so great that
proprietary grid generation software exists~\citep{GridPro:web}.

Overset mesh, or multipatch, techniques have started to make
an appearance in the numerical relativity community~\citep{2006CQGra..23S.553S,PhysRevD.77.103015,PhysRevD.80.024027,LLAMA-paper}.
However, these mesh structure techniques come with several limitations.
First, all tensor quantities are usually defined in an underlying
global Cartesian basis~\citep{LLAMA-paper,LLAMA-code}
due to the complicated
behaviour under coordinate transformations for some of the evolved fields.
However, it is often beneficial to define the grid functions in curvi-linear coordinates
directly.
More significantly, the
mesh structures are static in time~\citep{Clough_2015,2015CSE....17b..53S}. Finally, the current
state-of-the-art multipatch scheme in numerical relativity (Llama~\citep{LLAMA-paper,LLAMA-code})
can only solve the equations of hydrodynamics coupled to numerical relativity,
severely limiting its applicability for SMBBH accretion which requires
magnetic stresses to accurately model accretion processes.

Another method of modeling multiscale/multiframe is the use of Eulerian moving meshes
in hydrodynamics~\citep{arepo,Duffell_2011} and MHD~\citep{Duffell_2016}.
Alternatively, Arbitrary Lagrangian Eulerian (ALE) methods on unstructured meshes~\citep{PIERCE20053127,KENAMOND2014154}
seek to combine the strengths of Eulerian and Lagrangian techniques.
ALE algorithms allow the mesh to track the dynamics of the flow, 
reducing diffusivity, and have even been extended to include
MHD~\citep{RIEBEN2007534,2013JCoPh.247....1L,2014IJNMF..76..737B}. However, 
they encounter significant difficulty in the
presence of turbulent or vortex-like flows where meshes can easily tangle.
ALE methods handle this by mesh relaxation and remapping, but in doing so
lose much of their benefit whilst still incurring the cost of unstructured meshes.
Finally, yet other methods such as finite element discontinuous Galerkin, have shown significant
promise in their application to wave-like systems~\citep{Miller_2016,CHAN201822} and MHD~\citep{KARAMIHALASHI2016258,KIDDER201784}.

The \pw  infrastructure seeks to adopt several pieces of these approaches
into a single framework. Namely, the infrastructure is designed to
employ an overset mesh multipatch scheme where each mesh, or \enquote{patch},
is free to move dynamically in time with arbitrary mesh
refinement ratios. Furthermore, the patches may be freely defined in
any arbitrary coordinate basis or topology. In general, the patches
are also assumed to be unstructured/irregular with respect to one another.
Thus, the \pw infrastructure allows the deployment of user-specified mesh
refinement and a globally irregular mesh geometry whilst avoiding the
pitfalls of moving mesh methods.

This is accomplished through the MultiProgram-MultiData (MPMD) paradigm~\citep{MPMD}.
Each patch is itself a separate executable which communicates boundary
conditions in conjunction with all other patches using a client-router-server model
of MPI communication (for full details see Section~\ref{sec:BC}). Because each patch
is a separate executable, one may in principle turn on/off relevant physics packages
for time-dependent simulation regions. Examples of other simulation infrastructures
employing such a capability include combining fluid and molecular dynamics~\citep{PhysRevLett.96.134501} 
and modeling blood flow through the brain~\citep{GRINBERG2013131}.

The critical components missing
for our target science problem, multimethod abilities and higher-order methods,
are implemented for this study. In the next section, we motivate and further discuss extending \pw
to include multimethod abilities through the application of a proof-of-principle
toy model. Additionally, we have further improved core algorithms in the original  
\pw framework (in parallel to the \pwmhd development\citep{PWMHD}). These improvements have increased the global convergence order and
accuracy of the interpatch boundary conditions.

\subsection{Expanding the \pw Infrastructure}
The original \pw code was designed for hydrodynamic simulations.
As such, there exist optimization and algorithmic design choices
that make the \pw infrastructure incompatible with high order finite difference
methods. Namely, the original \pw code assumed a second order accurate method
of lines time integration, allowing each substep to be treated as an independent
Cauchy problem. Here we extend the \pw framework to be compatible with arbitrary 
time integration methods. This new version of \pw has the added benefit of removing
an algorithmic design limitation in which interpatch interpolation would use
previously interpolated data, thus increasing the accuracy. 

Furthermore, in order to extend the \pw infrastructure to ultimately 
simulate SMBBH mergers, we have implemented a new capability for \pwWave to handle
multiple state vectors simultaneously. 
Each state vector contains its own transformation and interpatch interpolation procedures, thus allowing the \pwWave 
code to, in principle, evolve two state vectors using a \textit{multimethod} approach
(such as updating the spacetime evolution using finite difference methods
and the equations of MHD using finite volume methods). However, we emphasize these
modifications are not restricted to finite difference or finite volume methods.

One of the most widely used evolution schemes for the spacetime in numerical
relativity is the so-called Baumgarte-Shapiro-Shibata-Nakamura (BSSN)
formulation~\citep{Shibata1995,Baumgarte1999}.
It is a reformulation of the Arnowitt-Deser-Misner (ADM) formulation of Einstein's
equations~\citep{Dirac1949,Arnowitt1962,Arnowitt2008}, and, like the ADM formulation,
adopts a $3+1$ foliation of spacetime~\citep{Darmois1927}. Unlike the ADM formulation
it introduces a conformal-traceless decomposition and conformal connection
functions (see~\citep{Baumgarte2010} for a textbook introduction).
The BSSN equations are a large set of non-linear, coupled partial differential equations (PDEs).

Provided the extensive complexities in implementing numerical relativity
in curvi-linear coordinates~\citep{Montero2012b,Baumgarte2013,Mewes18,Ruchlin:2018com},
which involves a reference-metric formulation of the BSSN 
equations~\citep{Bonazzola2004,Brown2009b,Gourgoulhon2012,Montero2012b}, 
and the large costs of such simulations, we first implement a proof-of-principle infrastructure here.
In an oversimplified sense, the equations of BSSN are a non-linear
wave-like system with stability criteria similar to a second order
formulation of the wave equation. Therefore, it is reasonable to first
demonstrate the new \pw infrastructure can evolve a scalar wave 
toy-model before proceeding to numerical relativity (which in and of itself
has required the deployment of large, open source collaborations for many years~\citep{ETK}).

In this paper, we derive and implement a fully coordinate invariant second order
formulation of the Klein-Gordon scalar wave equation into the \harm code. We
employ a non-traditional split of space and time which allows convenient
use of the full four dimensional manifold geometries already present in \harm.
The flexibility of this implementation is demonstrated by evolving the equations
on a time-dependent, dynamically warped mesh~\citep{WARPED}.
Furthermore, we fully couple this implementation to the \pw infrastructure
and demonstrate its efficacy using many arbitrary coordinate topologies and patch motions.

While our modifications to the \pw infrastructure allow for any arbitrary order 
discretization of the equations of motion in space and time, we elect to
use 4th order accurate methods for simplicity. The convergence order of
our interpatch interpolation and reflection-damping dissipation algorithms
are chosen to preserve the desired 4th order accuracy of our
discretization.

The global multipatch scheme is shown to be 4th order accurate with
dominant error sources coming from grid discretization, relatively
insensitive to the patch trajectories. Furthermore, we demonstrate
the extreme flexibility of our multipatch scheme by simulating all 
currently implemented
coordinate toplogies (Cartesian, cylindrical, and spherical) simultaneously.
These patches are also shown to be capable of moving independently in
rotating and linearly translating frames. We emphasize that these frames were chosen
for convenience, but any well-defined coordinate mapping to a new frame may
be used. Our tests are performed in full 3D and the error as the evolution proceeds is found
to always be continuous across mesh refinement, signifying that the interpatch
error is subdominant to the grid discretization error.

The rest of the paper is laid out as follows.
In Section~\ref{sec:code} we provide a full description of our \pwWave code, including 
relevant algorithmic details of the enhanced \pw code \cite{PWMHD} that are shared with \pwmhd. 
This includes the equations of motion, algorithms, and full interpatch
boundary condition procedure. In Section~\ref{sec:tests} we present
a suite of tests aimed at demonstrating the accuracy and flexibility
of our scheme. Finally, in Section~\ref{sec:conclusions} we provide
concluding remarks.

\section{Code Details}
\label{sec:code}
\subsection{Overview}

The \pw infrastructure is designed to accommodate arbitrary, time-dependent, moving
meshes, or \enquote{patches}, which together constitute a global problem. These individual
patches use separate executables which
are responsible for the physics occuring in their respective subdomains.
Patches are
broken into two categories: a unique \enquote{Global Patch} and an arbitrary number of
\enquote{Local Patches}. The local patches live on top of the global patch and their evolution
takes priority over global patch in regions of common coverage.

\begin{figure}[htb]
  \begin{center}
  \includegraphics[width=.5\columnwidth]{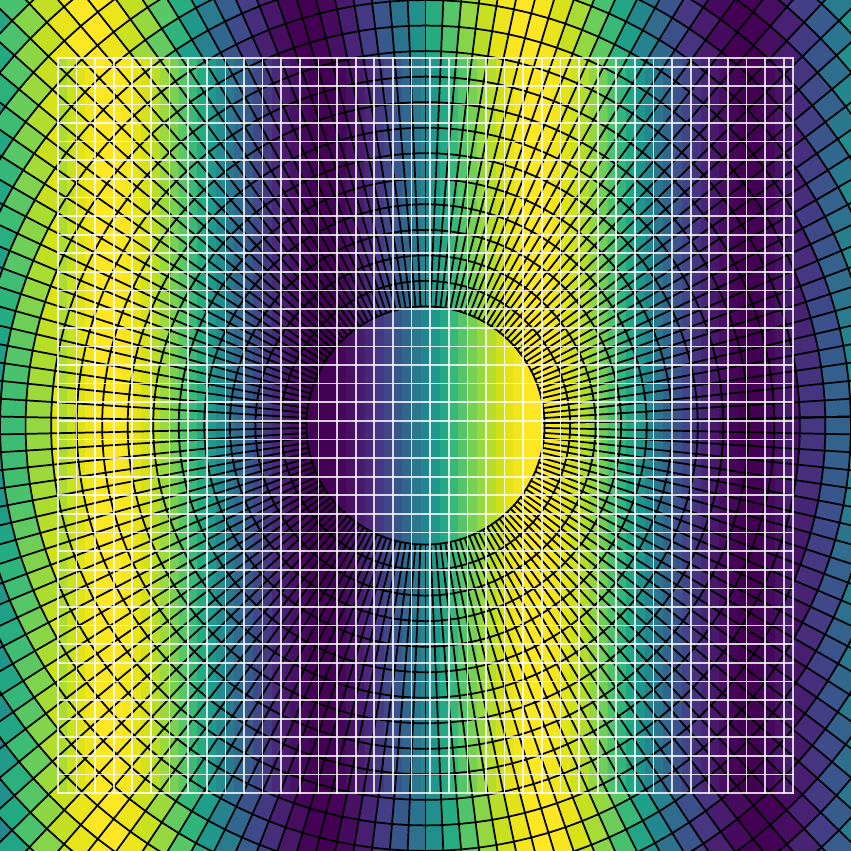}
  \end{center}
  \caption{Equatorial slice of a wave state through the spherical global
    patch with a local Cartesian patch covering the origin. Black lines show the
    spherical grid, while white lines denote the Cartesian local patch, 
    where every other grid line is shown.}
  \label{fig:patch_example}
\end{figure}

To provide intuition into the problem at hand, we plot a scalar plane wave for the simple
two patch configuration shown in Figure~\ref{fig:patch_example}. While both patches evolve
the same scalar plane wave, the coordinate representations and resolutions in which 
they do so are different. The global patch evolves in a discretization of spherical
coordinates which excises the coordinate singularities at the origin and the pole. Meanwhile,
the radially excised region is still included in the global domain by the inclusion of a Cartesian local patch.
The patch executables run concurrently under the MPMD paradigm and 
are connected through MPI communication of boundary condition data as necessary (see Section~\ref{sec:BC} for
full details).

Throughout the paper, unless otherwise noted, we use
units in which the speed of light $c=1$.  When used as tensorial indices,
we reserve Greek letters (e.g., $\alpha, \beta, \gamma, \ldots$) for
four-dimensional spacetime indices and Roman letters (e.g., $i, j, k, \ldots$) as
indices spanning spatial dimensions. We adopt the metric sign convention
of $\left(-, +, +, + \right)$ and the first index is coordinate time ($t$).
We follow the Einstein summation convention
where repeated upper and lower indices imply a summation.
However, the index $t$ does not obey this summation convention
and instead simply represents the index associated with coordinate time after 
splitting space and time.
\subsection{Equations of Motion}
\label{sec:eom}
We start with the standard scalar wave equation
\begin{equation}
  \nabla^2 \phi - \frac{1}{v^2} \partial^2_t \phi = S,
\end{equation}
where $S$ is some scalar source term.
Next, because our multipatch framework is designed to handle arbitrary coordinate systems
in arbitrary frames, we recast this system to a fully covariant expression. The 
d'Alembertian operator, $\nabla^2 - \frac{1}{v^2} \partial^2_t$, can be
expressed as
\begin{equation}
  g^{\mu \nu} \nabla_{\mu} \nabla_{\nu} \phi = S,
  \label{eq:covariant_wave}
\end{equation}
where $g^{\mu \nu}$ is the inverse metric tensor and $\nabla_{\mu}$ is the covariant derivative with respect to the metric $g_{\mu \nu}$ and
coordinates $x^{\mu}$, and we set $v=c=1$. For flat spacetime and Cartesian coordinates
the metric tensor and covariant derivative correspond to a $4\times4$ matrix $\textrm{diag}\left(-1,1,1,1\right)$ and the
partial derivative respectively. In~\ref{appendix:tensor_calc}
we provide a brief overview of the necessary differential geometry to understand this expression
and the subsequent ones for readers unfamiliar with this notation.

We can next expand this expression in terms of only partial derivatives of the scalar wave,
metric components, and Christoffel symbols of the second kind $\left({\Gamma^\alpha}_{\mu \nu}\right)$ as
\begin{eqnarray}
  S &=& g^{\mu \nu} \nabla_{\mu} \partial_{\nu} \phi \nonumber\\
  S &=& g^{\mu \nu} \left( \partial_\mu \partial_\nu \phi - {\Gamma^{\lambda}}_{\mu \nu} \partial_{\lambda} \phi \right).
\end{eqnarray}
The Christoffel symbols are themselves functions of derivatives of the metric tensor
\begin{equation}
  {\Gamma^\alpha}_{\mu \nu} = \frac{1}{2}g^{\alpha \lambda}\left( \partial_\mu g_{\lambda \nu} + \partial_\nu g_{\mu \lambda} - \partial_\lambda g_{\mu \nu} \right).
\end{equation}
We now define a coordinate time $t$ that is common among all patches and explicitly split
space and time.
\begin{eqnarray}
  S &=& g^{\mu t}\partial_{\mu}\partial_t \phi + g^{\mu i}\partial_{\mu}\partial_i \phi - g^{\mu \nu} \left({\Gamma^t}_{\mu \nu} \partial_t \phi + {\Gamma^j}_{\mu \nu} \partial_j \phi\right),\nonumber\\
  S &=& g^{tt}\partial_t\partial_t \phi + 2g^{t j}\partial_j \partial_t\phi + g^{ij}\partial_i \partial_j \phi 
     -g^{\mu \nu} \left( {\Gamma}^{t}_{\mu \nu} \partial_t \phi + {\Gamma^j}_{\mu \nu} \partial_j \phi \right).
\end{eqnarray}
Defining a new evolved quantity, $\Pi_t \equiv \partial_t \phi$, the expressions can be further written as
\begin{eqnarray}
  S &=& g^{tt} \partial_t \Pi_t + 2g^{tj}\partial_j \Pi_t + g^{ij}\partial_i \partial_j \phi 
     - g^{\mu \nu}\left({\Gamma^t}_{\mu \nu}\Pi_t + {\Gamma^j}_{\mu \nu} \partial_j \phi \right).
\end{eqnarray}

Our interpatch communication requires the interpolation of grid functions in different coordinate
systems and reference frames for arbitrary time-dependent patch configurations. We therefore will
need to coordinate transform $\Pi_t$ between patches. We define the one-form $\Pi_{\mu}$, of which $\Pi_t$ is a component,
as $\Pi_{\mu} = \partial_\mu \phi$. $\Pi_{\mu}$ obeys the standard coordinate transformation
\begin{equation}
  \Pi_{\mu^\prime} = \frac{\partial x^{\lambda}}{\partial x^{\mu^\prime}} \Pi_{\lambda}.
\end{equation}
We promote the spatial components $\Pi_j$ to evolved grid functions which are updated
through non-coupled differential equations (preserving the second order formulation).
Finally, we gather terms and simplify to obtain our complete equations of motion for the 
scalar wave in arbitrary coordinates, arbitrary reference frame, and any background spacetime as
\begin{eqnarray}
  \Pi_t &=& \partial_t \phi,\label{eq:eom1}\\
  \partial_t \Pi_t &=& \frac{1}{g^{tt}}\left[ S + g^{\mu \nu}\left({\Gamma^t}_{\mu \nu}\Pi_t + {\Gamma^j}_{\mu \nu} \partial_j \phi \right) 
              - 2g^{tj}\partial_j \Pi_t - g^{ij}\partial_i \partial_j \phi \right],\label{eq:eom2}\\
  \partial_t \Pi_i &=& \partial_i \Pi_t.\label{eq:eom3}             
\end{eqnarray}
We use the {\sc NRPy} code~\citep{Ruchlin:2018com,SENRNRPy:web} to generate the centered,
fourth order finite difference expressions for the right hand side (RHS) of the above evolution system.

In this framework, we may freely deploy any coordinate system or reference frame
without changing our code. We always finite difference in local coordinates and account
for curvature, motion, and curvilinear scale factors through a 4D coordinate transformation
acting on all tensorial quantities.
\subsection{Time Integration}
\label{sec:integration}
One may write an arbitrary order Runge-Kutta (RK) time integration step for
an equation of the form (see ~\cite{LLAMA-paper})
\begin{equation}
  \partial_t u\left(t,x^{i}\right) = f\left(t,x^{i}\right),
\end{equation}
discretized in time by $\Delta t$, as
\begin{eqnarray}
  u_{n+1} &=& u_n + \Delta t \sum_{l=1}^{s}b_l k_l\\
  k_l &=& f\left(t_n + c_l \Delta t, u_n + \Delta t \sum_{m=1}^s a_{lm}k_m\right).
\end{eqnarray}
Here, $u\left(t,x^i\right)$ is our wave state vector
$u = \left[\phi, \Pi_{\mu}\right]^T$ and $f\left(t,x^i\right)$
are their associated right hand sides in Equations~(\ref{eq:eom1}-\ref{eq:eom3}).
We elect to use the ``classic'' fourth order method~\citep{1992nrca.book.....P}
and provide the Butcher tableau in Table~\ref{tab:rk4}.
\begin{table}[htb]
\centering
\begin{tabular}{c | c c c c}
  \centering
 0   & 0   & 0   & 0 & 0\\ 
 1/2 & 1/2 & 0   & 0 & 0\\
 1/2 & 0   & 1/2 & 0 & 0\\
 1   & 0   & 0   & 1 & 0\\
 \hline
 & 1/6 & 1/3 & 1/3 & 1/6
\end{tabular}
\caption{Butcher tableau for our fourth order Runge-Kutta. The
  coefficients are $b_l$ (bottom line), $c_l$ (first vertical
  column), and $a_{lm}$ in the remainder of the table.}
\label{tab:rk4}
\end{table}

Recall that this code was ultimately designed to update the
Einstein Field Equations (EFE) coupled to MHD.
When integrating wave-like systems with the method of lines, one either needs 
to use partially implicit RK methods or higher than second order fully explicit RK
methods, even in Cartesian coordinates~\citep{CorderoCarrion:2012ic,Cordero-Carrion:2014}. 
However, finite volume implementations of GRMHD are usually only
formally second order accurate (due to the approximation of volume integrals by the
midpoint rule) and furthermore drop to first order accuracy at shocks. It would 
be computationally wasteful to integrate the MHD equations at higher than second
order in time.

We therefore
write our time integration of the wave into two steps, compatible
with a future mixed order integration scheme. In this scheme
we first update the state vector to the half-step and return
that result. We then update to the full-step.
More explicitly, we update our initial wave state vector
$u_n$ to $u_{n + 1/2}$ as
\begin{eqnarray}
  K_1 &=& f\left(t,u_n\right)\label{eq:k1}\\
  K_2 &=& f\left(t+c_2\Delta t, u_n +  a_{21} K_1\Delta t\right)\label{eq:k2}\\
  K_3 &=& f\left(t+c_3\Delta t, u_n + a_{32} K_2\Delta t\right)\label{eq:k3}\\
  u_{n + 1/2} &=& u_n + \Delta t\left( b_1 K_1 + b_2 K_2 + b_3 K_3\right).
\end{eqnarray}
We then update $u_{n+1/2}$ to $u_{n+1}$ as
\begin{eqnarray}
  K_4 &=& f\left(t + c_4\Delta t, u_n + a_{43} K_3\Delta t\right)\label{eq:k4}\\
  u_{n+1} &=& u_{n+1/2} + \Delta t \left(b_4 K_4\right).
\end{eqnarray}

The full description of the boundary conditions we apply 
before calculating the slope estimates can be found in Section~\ref{sec:BC}.
To reduce high frequency noise arising from reflections at mesh refinement boundaries, 
we add Kreiss-Oliger dissipation~\citep{Kreiss73} to the RHS of all evolved variables. 
An n-th order dissipation operation can be written as (see e.g.~\citep{Alcubierre2008,ETK-KO}):
\begin{equation}
  f\left(t,\vec{x}\right) =  f\left(t,\vec{x}\right) + \left(-1\right)^{\left(n+3\right)/2} \epsilon \frac{1}{2^{n+1}} dx^i \partial_i^{n+1} u\left(t,\vec{x}\right), 
\end{equation}
where $\epsilon$ is the dissipation strength coefficient (set to 0.005 for this paper), and $dx^i$ is the
cell spacing in the evenly sampled, logically Cartesian, numerical grid. To preserve the
fourth order convergence of our time integration and spatial finite difference stencils,
we set $n=5$ and arrive at the following addition to the RHS:
\begin{equation}
  f\left(t,\vec{x}\right) =  f\left(t,\vec{x}\right) + \epsilon \frac{1}{2^{6}} dx^i \partial_i^{6} u\left(t,\vec{x}\right).
\end{equation}

\subsection{Boundary Conditions}
\label{sec:BC}
\begin{figure}[h!tb]
  \centering
  \includegraphics[width=0.93\textwidth]{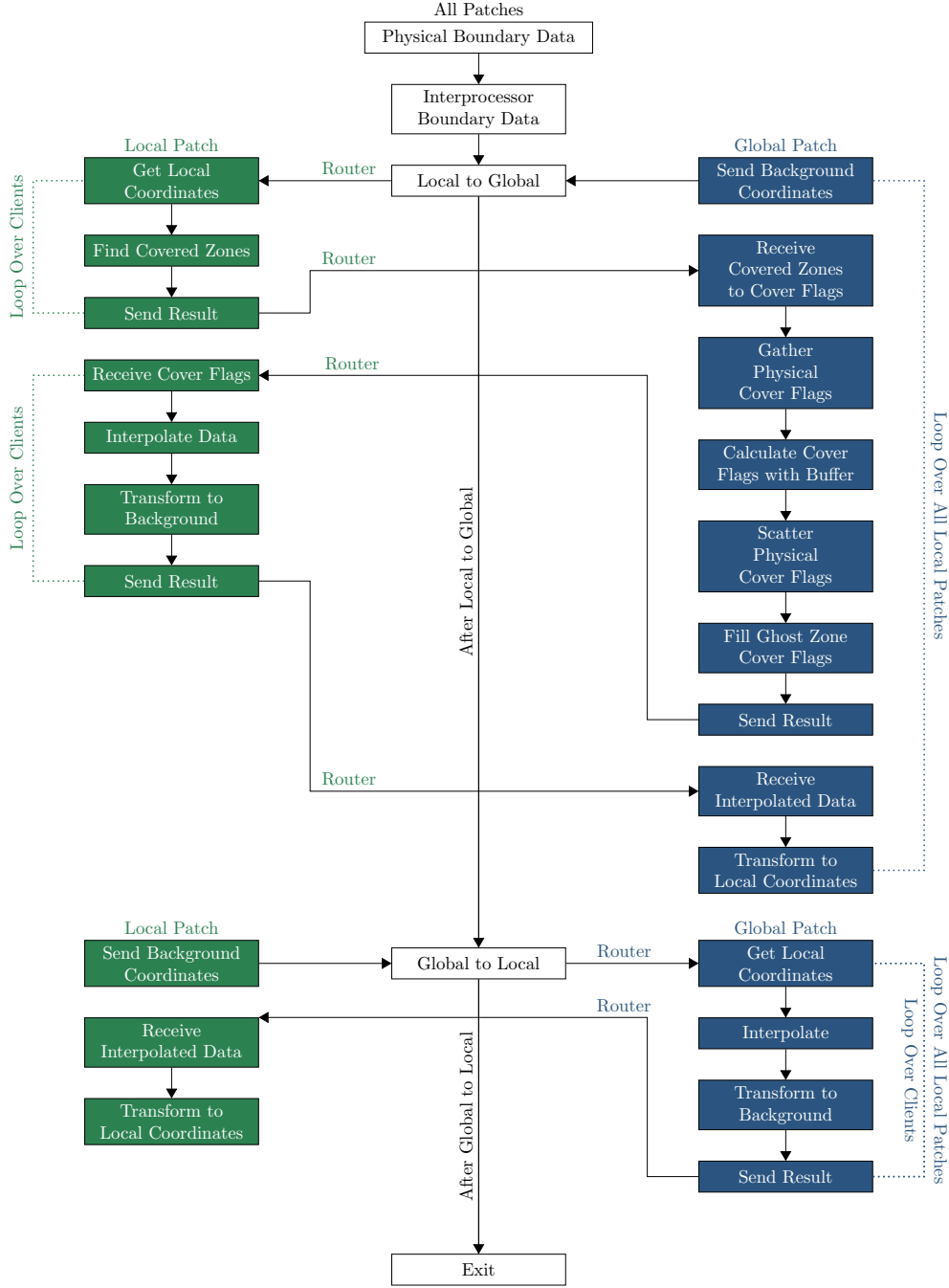}
  \caption{Flow chart for the hierarchy of boundary conditions for our code. A \enquote{client} is
  an individual CPU domain that lives on another patch for which this CPU is responsible. 
  Operations in the left column are
  performed by local patch CPUs (colored green), the middle column by all CPUs (no color), and the right column
  by global patch CPUs (colored blue). 
  We denote interpatch communication which goes through a router CPU by placing a color coded ``Router'' label. The color
    coding specifies which patch the router CPU is a member of.}
  \label{fig:BC_order}
\end{figure}

A critical component of our infrastructure is the hierarchy of
boundary conditions that we employ on the slope estimates $K_{1-4}$
in Equations~\ref{eq:k1}-\ref{eq:k3} and \ref{eq:k4}.
A version of this hierarchy was developed as part of the enhanced \pw code (see \cite{PWMHD} for details), but several further modifications
were necessary to facilitate arbitrary order time integration  
and further customization to our problem.
We provide a diagram of our boundary condition procedure in Figure~\ref{fig:BC_order}.

From Figure~\ref{fig:BC_order}
we see that our boundary condition priority order is
interpatch boundary data over interprocessor/periodic boundary data over
simulation domain boundary data (analytic for our tests). We therefore lay down our boundary
data into the ghost zones in the inverse of this priority order, 
\begin{equation}
  \text{Boundary Data} \to \text{Interprocessor Data} \to \text{Interpatch Data},\nonumber
\end{equation}
allowing interpatch
boundary data to replace existing boundary conditions. By successively rewriting
boundary data in this order we ensure BCs are consistent with the evolution. 

Interpatch boundary conditions are applied using the \pw framework.
\pw provides the ability for various local meshes, or patches, of arbitrary
resolution and coordinate topology to move freely over an underlying mesh
or global patch (recall our example in Figure~\ref{fig:patch_example}).
This facilitates the ability to construct local patches which are better
suited to capture the physics in the encompassed area.
As such, we deactivate cells on the global
patch which are covered by local patches.  This maintains consistency between
the two patches in the overlapping region and eliminates redundant effort on the
global patch.

However, the cells immediately adjacent to the local patch which are evolved
on global patch require boundary conditions. These effective ghost
zones are populated via interpatch interpolation from data which resides on the
local patch. Similarly, the physically evolved cells on the edge of local patch
require boundary conditions. These ghost zones are filled by interpatch
interpolation of data which resides on the global patch. We refer to these
two methods of providing boundary conditions as Local to Global and Global to Local
(see Sections~\ref{sec:local_to_global}~and~\ref{sec:global_to_local},
respectively).

Interpatch data is always interpolated in the local, regular coordinate basis of
the patch on
which the source data resides. We apply centered fifth-order Lagrange polynomial interpolation.
In 1D Lagrange polynomial interpolation can be written as
\begin{equation}
  u(x) = \sum_{i=1}^{n+1} P_i(x),
\end{equation}
where $u$ is the function being interpolated at location $x$,
$n$ is the order of interpolation, and
\begin{equation}
  P_i(x) = u(x_i)\prod_{l=1, l\neq i}^{n+1} \frac{x-x_l}{x_i-x_l}.
\end{equation}
Here $u(x_i)$ is the function to be interpolated evaluated at the discrete point $x_i$
and sampled at points $x_1$ to $x_{n+1}$.

In practice, our 3D interpolation can be written as the tensor product of each
dimension's individual interpolation as
\begin{equation}
  u(x,y,z) = \sum_{i=i_c - 2}^{i_c+3}\sum_{j=j_c - 2}^{j_c+3} \sum_{k=k_c - 2}^{k_c+3}u_{i,j,k}P_{i}^{j,k}(x)P_{j}^{i,k}(y)P_{k}^{i,j}(z)
\end{equation}
where
\begin{equation}
  \label{eq:Pijk}
  P_{i}^{j,k}(x) = \prod_{l=i_c - 2, l \neq i}^{i_c + 3} \frac{x - x_{l,j,k}}{x_{i,j,k} - x_{l,j,k}}.
\end{equation}

We use indexing $u_{i,j,k}$ and $x_{i,j,k}$ to represent the discrete grid functions
  and coordinates, respectively.  The $\left(i_c,j_c,k_c\right)$ indices are the closest discrete locations to 
  the point of interpolation without exceeding it in any dimension, i.e.  $\left(x,y,z\right)$
  lies  between $\left(x,y,z\right)_{i_c,j_c,k_c}$ and
  $\left(x,y,z\right)_{i_c+1,j_c+1,k_c+1}$. In Equation~\ref{eq:Pijk},
  we denote which index/dimension is iterated over in the product operation with the index $l$ by placing that dimension's index as a subscript to $P$.

\subsubsection{Global to Local}
\label{sec:global_to_local}
\begin{figure}
  \centering
  \includegraphics[width=0.45\textwidth]{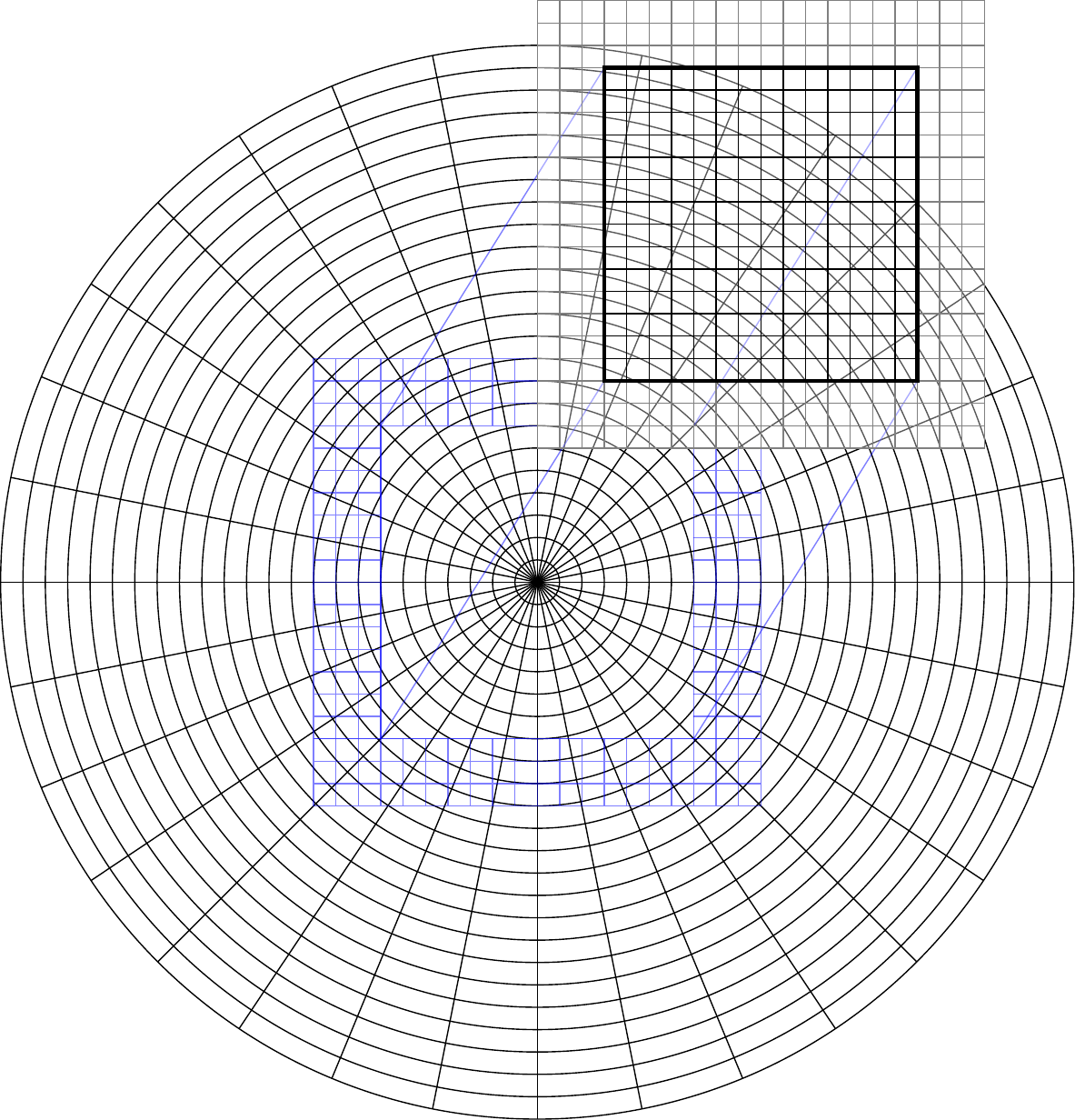}
  \caption{Schematic diagram of Global to Local interpolation zones in the
  equatorial plane of a spherical global patch. The local Cartesian patch,
  which actually resides over the origin, is offset for visualization. The 
  solid black line denotes the edge of the physical cells on the local Cartesian
  mesh and gray gridlines denote physical ghost zones requiring interpatch data.
  The locations of these ghost zones on the global patch are marked by blue 
  grid lines on the spherical mesh.}
  \label{fig:global_to_local}
\end{figure}
Although in Figure~\ref{fig:BC_order} we see that our algorithm first performs
Local to Global, we begin our discussion with the simpler Global to Local.
In the absence of overlapping local patches, the order of these BC operations
is interchangeable. However, we emphasize that this interchangeability is a 
feature unique to the \pwWave version of \pw.

As a local patch traverses the global patch, the edges of that local patch
require boundary conditions. Furthermore, because \pw allows
dynamic patches and different coordinate topologies, coordinate locations
are generally irregular with respect to the global patch mesh
where the required data resides. We schematically visualize this in Figure~\ref{fig:global_to_local}
for the two patch configuration of Figure~\ref{fig:patch_example}. 

Consider a configuration with one global patch and some arbitrary number
of local patches. Each CPU evolves in a numeric coordinate basis,
${X_{j}^\alpha}^{\prime}$. Here j refers to the patch ID, runs from $0$ to $\rm{NPATCHES} - 1$,
and global patch is denoted by $j=0$. 
Furthermore, each patch member
knows the functional mapping from its numeric coordinate basis to a physical coordinate
basis of the same coordinate topology, ${X_{j}^\alpha}\left({X_j^\alpha}^\prime\right)$,
where $X_{j}^\alpha$ is either Cartesian, cylindrical, or spherical coordinates.
Because each $X_j^\alpha$ is in principle irregular with respect to one another, and not known globally by all patches,
all interpatch communication goes
through a common background Cartesian coordinate system,
$X_B^\alpha$.

\fbox{Send Background Coordinates $\to$ Get Local Coordinates}
  \footnote{We use boxed text to denote where each portion
    of our algorithmic discussion is located in Figure~\ref{fig:BC_order}}.
In Local to Global, each local patch processor transmits its ghost zone
locations, denoted by gray and blue cells in Figure~\ref{fig:global_to_local},
to the global patch CPUs (via a router - see Section~\ref{sec:parallelism}
  for discussion on parallelism)
in the common background Cartesian coordinates: $X_B^\alpha \left(X_j^\alpha\right)$. 
Upon receiving these coordinate locations, the global patch processor
transforms the points from background Cartesian
to its own local numerical coordinate basis:
${X_0^\alpha}^{\prime}\left(X_0^{\alpha}\left(X_B^{\alpha}\right)\right)$.

\fbox{Interpolate $\to$ Transform to Background $\to$ Send Result}
Global patch CPUs then attempt to locate and interpolate the state
vector at each point in the ${X_{0}^\alpha}^\prime$
coordinate basis. If the data is successfully interpolated, the state vector is then transformed
on the global patch processor to the background coordinates,
\begin{equation}
  {X_0^\alpha}^\prime \to {X_0^\alpha} \to X_B^\beta,\nonumber
\end{equation}
 and then transmitted to local patch ``$j$''. 
Throughout the paper, we adopt the notation $X_j^\alpha \to X_k^\beta$
as representing the operation of mapping the state vector from the $X_j^\alpha$
coordinate basis to the $X_k^\beta$ coordinate basis.

\fbox{Receive Interpolated Data $\to$ Transform to Local Coordinates}
The Local patch processor then transforms the interpolated data 
from the background coordinates
to the local patch numerical coordinate basis:
\begin{equation}
  X_B^\alpha \to X_j^\alpha \to {X_j^\alpha}^\prime.\nonumber
\end{equation}

If a global patch processor finds that
it cannot succesfully interpolate to a point, 
the state vector is marked
to be skipped at that point. Upon receiving the state vector on the local patch processor,
 only those state vectors successfully interpolated are used.
Referring to Figure~\ref{fig:BC_order} this amounts to assuming that the point
should be filled already via either physical boundary conditions or intrapatch interprocessor boundary
conditions.

\subsubsection{Local to Global}
\label{sec:local_to_global}
\begin{figure}[htb]
  \includegraphics[width=.5\columnwidth]{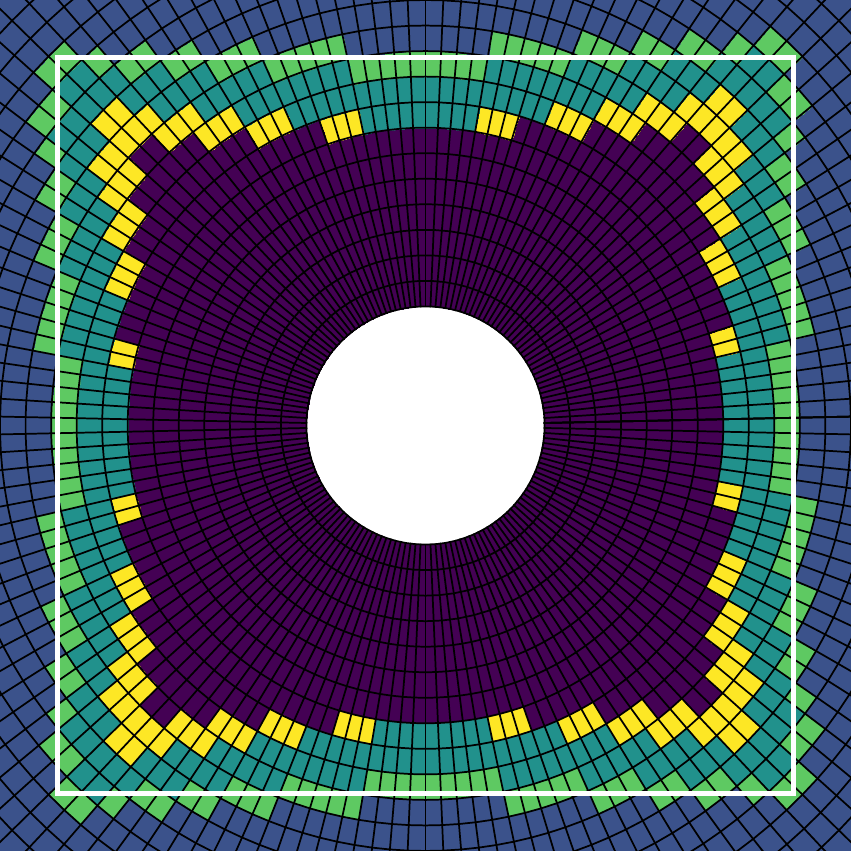}\includegraphics[width=.5\columnwidth]{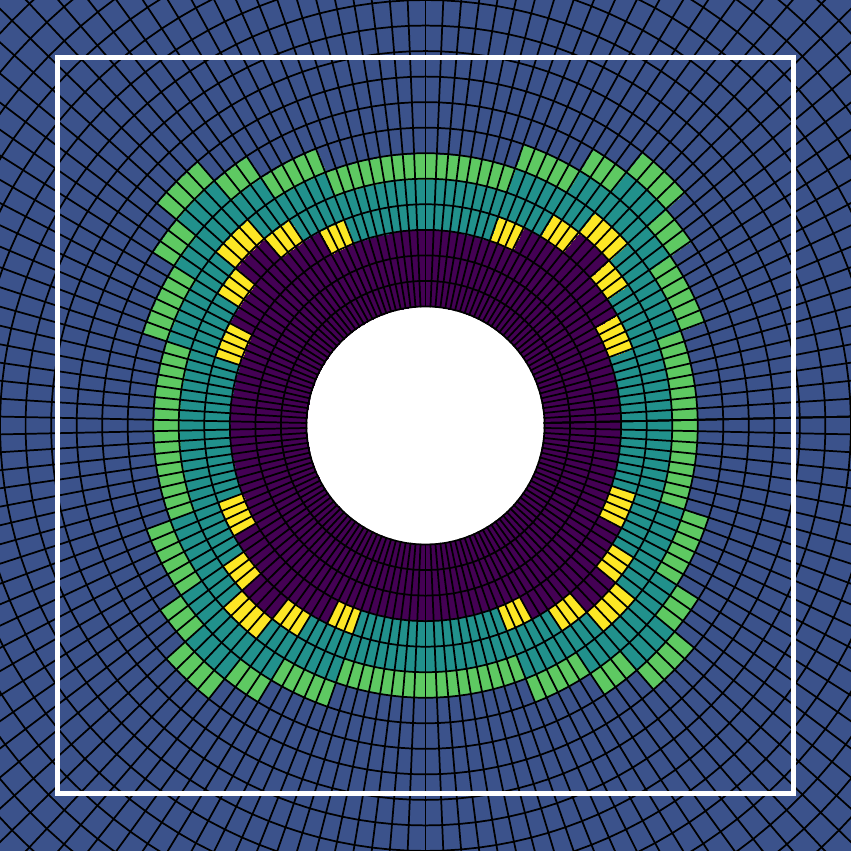}
  \caption{Example interpolation flag configuration for a Cartesian
    local patch over the origin of a global spherical patch
    in the equatorial plane. (Left) We plot the original flag implementation
    in~\cite{PATCHWORK}. (Right) We plot the new implementation with a buffer
    width of 4 cells. Purple zones are deactivated and play no role in the
    cell update, blue zones are evolved cells, and all other zones are
    flagged for interpatch interpolation. The white box denotes the edges
    of the local patch. Green cells are those which are within the ghost
      zone depth along any coordinate direction from evolved cells. Yellow cells are those within
      a ghost zone depth diagonally from evolved cells.}
  \label{fig:cflag_example}
\end{figure}
The other interpatch boundary condition, Local to Global, is responsible
for determining which cells may be deactivated on the global patch domain as well
as populating the data in the effective ghost zones underneath the local patches.
This algorithm and its parallelization (see Section~\ref{sec:parallelism}),
represents the most significant departures from the original \pw 
framework.

Consider again a configuration with
arbitrary number of local patches. In this scenario, we assume that no local patches
overlap and can therefore treat each local patch's interpatch BC independently. 
Below we describe the BC which is performed for each local patch individually.

\fbox{Send Background Coordinates $\to$ Get Local Coordinates
$\to$ Find Covered Zones $\to$ Send Result} 
At the start of each timestep, the global patch processor must first ascertain
which cells are covered by a local patch. 
For a given local patch $j$, the global patch processor
first sends its coordinates in the background
Cartesian frame, $X_B^\alpha\left(X_0^\alpha\right)$, to a local patch processor.
The local patch
processor then converts the coordinates to its own numeric coordinate basis
${X_j^\alpha}^\prime\left(X_j^\alpha\left(X_B^\alpha\right)\right)$.
It may then ascertain which global patch cells
the local patch covers,
and send the result back to the global patch processor.

The reason we calculate which cells are covered on the local patch processor is because
the global patch cells are generally irregular with respect to the local patch cells.
Therefore, one cannot simply ask if a cell is within the bounds of the local patch
coordinates. However, the local patch numerical coordinates, ${X_j^\alpha}^\prime$,
are required by \harm
to be regular and each global patch point, ${X_0^\alpha}$, may be determined to be
within/exterior to the bounds of those regular local patch numerical coordinates.

\fbox{Receive Covered Zones to Cover Flags}
Upon receiving the determination of which global patch cells are covered, the
original \pw code would immediately calculate which covered cells are within an update stencil
width of an uncovered cell. These global patch cells would then be flagged for interpatch
interpolation. We plot the interpolation flag markers (cover flags) for our two patch example using this
original algorithm in the left frame of Figure~\ref{fig:cflag_example}.

\fbox{Calculate Cover Flags with Buffer}
However, in extending to arbitrary order methods this implementation 
leads to an issue. Consider that after filling the effective ghost zones in the global patch
with data from the local patch, \pw must then call Global to Local. 
Looking at Figures~\ref{fig:global_to_local}~and~\ref{fig:cflag_example}, one may quickly observe
that the ghost zones of the local patch would be interpolated into using data previously interpolated
onto the global patch domain.
This would drop the convergence of our interpolation
accuracy, requiring higher order interpolation stencils and increased computational expense to overcome
(both via more expensive interpolation weights and an increase in the required ghost zone count).

To overcome this limitation, we assume there exists a region of covered zones on the global patch domain
capable of capturing
the physics in addition to local patch. This is well posed as generally we would avoid significant cell size
discrepancies at the patch boundaries to minimize numerical artifacts. In the right frame of Figure~\ref{fig:cflag_example}
we plot the same patch flag configuration including a ``buffer'' of evolved cells on global patch that reside
immediately under the edges of the local patch. By implementing
this buffer we wish to eliminate the interpolation of interpolated data present in the original \pw code~\citep{PATCHWORK}.

\fbox{Send Result $\to$ Receive Cover Flags $\to$ Interpolate Data}
Once the cells that will serve as effective ghost zones under the local patch are determined,
the global patch processor
sends their coordinate locations back to the local patch processor in the background Cartesian coordinates and
poisons the data of fully covered cells which may be deactivated. The local patch processor then transforms the 
effective ghost zones to the local numerical coordinate basis,
${X_j^\alpha}^\prime\left(X_j^\alpha\left(X_B^\alpha\right)\right)$,
and attempts to interpolate the state vector at each point in the ${X_j^\alpha}^\prime$ coordinate basis.

\fbox{Transform to Background $\to$ Send Result}
If the data is successfully interpolated, the state vector is then transformed on the local patch processor
to the background coordinates,
\begin{equation*}
  {X_j^\alpha}^\prime \to X_j^\alpha \to X_B^\alpha,
\end{equation*}
and then transmitted to the global patch processor.

\fbox{Receive Interpolated Data $\to$ Transform to Local Coordinates}
Upon receiving the interpolated data, the
global patch processor transforms the state vector to its own numerical coordinate basis:
\begin{equation*}
  X_B^\alpha \to X_0^\alpha \to {X_0^\alpha}^\prime.
\end{equation*}
Similar to Global to Local, if the point was found to be uninterpolatable,
then the global patch processor skips the point and assumes that it has been previously
filled by intrapatch boundary conditions.

A significant deviation of our time integration from the original \pw
and subsequent enhanced \pw and \pwmhd codes is that we use 4th order
methods. As such, we can no longer treat the
timestep as two
independent Cauchy problems of size $\Delta t/2$. This is because we require information about the state vector
at the half-step to fully update to $t+ \frac{1}{2}\Delta t$. To accommodate this, we alter the \pw 
implementation to keep constant the memory locations of live, covered, and effective ghost zones (cover flags)
on the global patch processors throughout the entire timestep.
The determination of which cells are live, covered, or effective ghost zones is done once the full timestep has been performed.

However, because we allow time-dependent transformations
between physical and numerical coordinates on each patch, we must update these transformations 
when coordinate time is changed at the substeps. 
Additionally, because our local patches are free to move on the global patch
in time, we also need to update the coordinate mappings between the global, local, and background coordinates
at the same substeps. Provided these considerations, the Local to Global interpatch boundary
condition at substeps is precisely that described above, but skipping the calculation of which
zones are covered, live, or effective ghost.

A choice required for the Local to Global interpatch boundary condition
is the specification of the buffer zone width. Our interpolation stencil extends at most
$NG=3$ cells away from any given point, where $NG$ is the number of ghost zones. 
We therefore specify the width of the buffer of live
cells underneath the local patch to $NG+1$ cells from the edges of local patch.
This allows for a local patch
to traverse one global patch cell during the timestep in any number of spatial dimensions.
Currently, it is encumbant upon the user to verify that the specified patch trajectories
do not violate this condition. However, the buffer width is a freely specifiable parameter that may also be
increased to relax this constraint.

Finally, special care must be taken for physical ghost zones (ghost cells which are not interprocessor ghost zones)
which are covered by a local
patch. In the case of our two patch example in Figure~\ref{fig:patch_example}, all
inner radial ghost zones are covered by the local patch. These zones may be immediately
adjacent to physically evolved cells, particularly in the buffer region, and will be used in
cell updates. Since there is no reason to assume that the spherical BCs of the
global patch match the dynamics on the Cartesian local patch, we mark all covered physical
ghost zones for interpatch interpolation. This ensures that our boundary conditions
are consistent with the local patch evolution, while zones which cannot be filled
by interpolation will preserve the intrapatch BCs of global patch by the skipping mechanism
previously described.

\subsubsection{Parallelism}
\label{sec:parallelism}
\begin{figure}[htb]
  \centering
  \includegraphics[width=.5\textwidth]{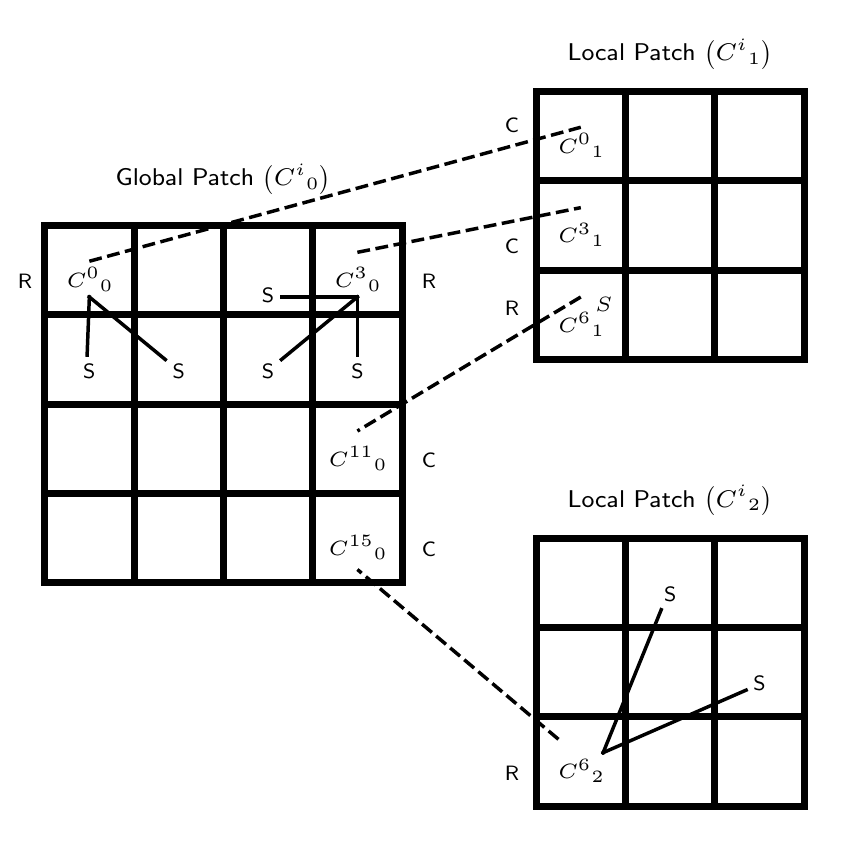}
  \caption{Reproduction of figure 3 from \cite{PATCHWORK} schematically demonstrating
  the client-router-server paradigm for 3 patches employing multiple routers. Grids represent
  CPU domain decompositions. CPUs are marked as a client (C), server (S), or router (R). Solid
  lines connecting a router CPU to a server CPU denote intrapatch communication and dashed
  lines connecting a client CPU to a router CPU denote interpatch communication. Each such line
  denotes communication out and back between the connected CPUs.}
  \label{fig:client-server-router}
\end{figure}

In addition to patches being run in parallel via different executables which
communicate BCs through MPI, each patch is itself MPI domain decomposed. We
therefore must distinguish between interpatch and intrapatch communication.
Furthermore, CPUs assigned to one patch do not know how the other patches 
are subdivided and assigned, and consequently do not know from which CPU to request data. 
\pw handles this transfer of data from an 
individual patch CPU to the appropriate CPU on another patch via
a client-router-server model (for a full discussion please see~\cite{PATCHWORK}). 
Important modifications were necessary to scale well to more than a few hundred MPI 
ranks, and details of these changes to the original \pw code will be reported in \cite{PWMHD}.

Let each CPU be denoted by ${C^i}_j$, where $i$ denotes the CPU number on a given patch
and $j$ denotes the patch ID. We note that we have swapped the index locations from those
in \cite{PATCHWORK} for consistency with our coordinate indexing in Sections~\ref{sec:global_to_local}~and~\ref{sec:local_to_global}.
Patch IDs run from 0 to $N_{\rm PATCHES} - 1$ 
and global patch is denoted with $j=0$. On any given patch, $i$ runs from 0 to ${N_{\rm CPU}}_{j} - 1$
where ${N_{\rm CPU}}_{j}$ is the number of CPUs on the $j^\mathrm{th}$ patch. 

Consider the simpler Global to Local interpatch BC. Each ghost zone on a local patch
will attempt to fill its data with interpolated data from the global patch. In this
instance, each ${C^i}_{j\neq 0}$ containing physical ghost zones becomes a ``client'' which requires data.
However, because the client CPU does not have any knowledge of which CPU on the other patch
contains the necessary data, it
sends a request to the global patch via a ``router'' CPU. This router can 
distribute the requests from the client CPU to the appropriate global patch
``server'' ${C^i}_0$ CPUs that contain the data. Finally, the server CPUs provide the
requested data and forward it back to the client CPU through the router CPU.
We demonstrate this process 
schematically in Figure~\ref{fig:client-server-router} (see also Figure 3 of \cite{PATCHWORK}).

Similarly, when data requests are forwarded between patches in Local to Global
it is transmitted from client CPUs requiring information to server CPUs containing the information
via router CPUs on the server's patch. This includes the sending of coordinates from global patch CPUs
to local patch CPUs to determine which zones are covered as well as the request for interpolated data
made by global patch CPUs to the local patch CPUs.

\fbox{Gather Physical Cover Flags $\to$ 
Calculate Cover Flags with Buffer $\to$ 
Scatter Physical Cover Flags}\\
Our Local to Global interpatch BC further differs from the original \pw implementation in the parallelization 
of the calculation of the effective ghost zone locations on global patch. In the original \pw
method, each ${C^i}_0$ would calculate its own interpolation flags and then send data requests
as necessary to the local patch routers at each substep. However, because our method employs
a buffer extending $NG+1$ cells, each ${C^i}_0$ cannot be assumed to know the location of all
patch edges required for calculating the interpolation flags. 

Therefore, at the start of each
timestep the global patch CPUs all forward the covered/uncovered status of their physical cells
to a single global patch CPU ${C^0}_0$. ${C^0}_0$ then calculates
the locations of effective ghost zones and covered states for all physical cells.
We refer to physical cells as the set of cells which are a part of the computational
domain, but are not a ghost cell.
This information is then forwarded back to all ${C^i}_0$ to populate both physical
and interprocessor ghost zone interpolation flags.

\section{Code Tests}
\label{sec:tests}

To test our code we solve the wave equation using as initial and boundary data a simple plane wave,
\begin{equation}
  \phi\left(t,x\right) = \sin\left[ \frac{2\pi}{L}\left(x - t\right)\right] + 2,
  \label{eq:wave_solution}
\end{equation}
where $L=20$ is a free parameter used to scale the period of the wave. The
addition of 2 removes zero crossings to simplify the normalization of error
calculations. 

While in Cartesian coordinates this is a 1D problem, in curvilinear coordinates
our entire covariant infrastructure is required to recover the correct solution.
This is because our code always finite differences with respect to the numerical
coordinate basis and uses the metric tensor and Christoffel symbols to generalize
the derivatives. Additionally, the one-form $\Pi_{\mu}$ is coordinate dependent
and therefore changes when solving in different coordinates/frames. Therefore, when
we run multiple patches in different coordinate topologies, the entire transformation
and interpolation infrastructure must also be used to correctly translate the solution
in the BCs. Although symmetries may be permitted by the solution, we always
run our multipatch tests in full 3D without making any symmetry assumptions.
As such, by running curvilinear patches with a Cartesian problem, we test
every component of our infrastructure simultaneously.

In all tests we apply analytic BCs at the physical edges of the global domain.
This allows us to configure our patches arbitrarily without concern to exposed 
edges or proper transmissive boundaries. Furthermore, because the BCs are analytic,
any reflections and errors introduces by the mesh refinements and evolution cannot
be transmitted out of the domain. In setting our tests up this way, we seek to explore
the ``worst case scenario'' where any mesh-to-mesh error/noise is unable to escape the
domain and more likely to disrupt the evolution than a production simulation employing
numeric/transmissive BCs.

Our timestep is always constant throughout the evolution because the numeric
cell crossing time for our wave characteristic is fixed in time. Finally,
we set the Kreiss-Oliger dissipation coefficient $\epsilon = 0.005$ for all
simulations. A small dissipation coefficient assures the damping
of high frequency noise while not affecting the physical solution.

\subsection{Test One}
\label{sec:warped}

\begin{table}
\centering
\begin{tabular}{l r }
\textbf{Parameter} & \textbf{Value}\\
\hline
$\delta_{x1}$, $\delta_{x2}$, $\delta_{x3}$, $\delta_{x4}$ & 0.1 \\
$\delta_{y1}$, $\delta_{y2}$, $\delta_{y3}$, $\delta_{y4}$ & 0.1 \\
$h_{x1}$, $h_{x2}$, $h_{x3}$, $h_{x4}$  & 20.0\\
$h_{y1}$, $h_{y2}$, $h_{y3}$, $h_{y4}$  & 20.0\\
$a_{x10}$, $a_{x20}$, $a_{y10}$, $a_{y20}$ & 1.0\\
$x_{\rm min}$, $y_{\rm min}$ & -20.0\\
$x_{\rm max}$, $y_{\rm max}$ & 20.0
\end{tabular}
\caption{Parameters of the warped Cartesian grid used for our simulation.
  The full expressions relating these parameters are in Equations 20 and 21
  of \cite{WARPED}.}
\label{tab:warped}
\end{table}
To begin, we test our wave solver independent of the \pw BC infrastructure. 
We evolve the plane wave solution of Equation~\ref{eq:wave_solution} in a time-dependent double fish-eye warped
Cartesian coordinate system~\citep{WARPED}. In this coordinate system there are two
regions of dynamic warping which orbit counter-clockwise at an orbital frequency
of $\Omega_{\rm warp} = \pi / 50$. We tabulate the parameters of the warped grid
in Table~\ref{tab:warped}.

For this study (and in our upcoming paper about \pwmhd \citep{PWMHD}), we implemented 
the infrastructure to map any spacetime known by \harm
from a background reference frame to some arbitrary time-dependent reference frame. 
The background frame here would correspond to $X_B^\alpha$
in the \pw infrastructure. Specifically, we pass the background coordinates
to the routines responsible for evaluating the metric, which then transform the resultant metric tensor via
\begin{equation*}
  X_B^\alpha \to X_{\rm rotating}^\alpha \to X_{\rm WARPED}^\alpha.
\end{equation*}

In addition to testing the dynamic coordinate and covariant infrastructure of
the wave solver, we wish to test this generalized frame infrastructure. 
To do this, we place the mesh into a 
frame rotating clockwise at a frequency of $\Omega_{\rm frame} = 0.01$
about the $z$ axis with respect to the background coordinates.
That is,
  \begin{equation}
    \frac{\partial}{\partial X_B^\alpha} X_{\rm rotating}^\alpha = \begin{bmatrix} \cos \left(\Omega t\right) & \sin \left(\Omega t\right) & 0 \\ -\sin \left(\Omega t\right) & \cos \left(\Omega t\right) & 0 \\ 0 & 0 & 1 \end{bmatrix},
  \end{equation}
  where $\Omega = \Omega_{\rm frame}$.  

As we are not employing the full interpatch BCs and only testing the enhanced \pw 
coordinate and \harm metric modifications in conjunction with our wave solver,
we run this test in 2D. Our grid contains $100 \times 100$
cells, producing a numerical cell size of $dx^1 = dx^2 = 0.01$. We set the timestep
to half this numerical cell size: $dt = 0.005$. The simulation domain extends from
$-20$ to $20$ in $x$ and $y$ in the local rotating frame. We normalize the wave
period of Equation~\ref{eq:wave_solution} by setting $L = 20$. This test
employs every component of the generalized coordinate/metric infrastructure within
the $x-y$ plane.

\begin{figure}[htb]
  \centering
  \includegraphics[width=\textwidth]{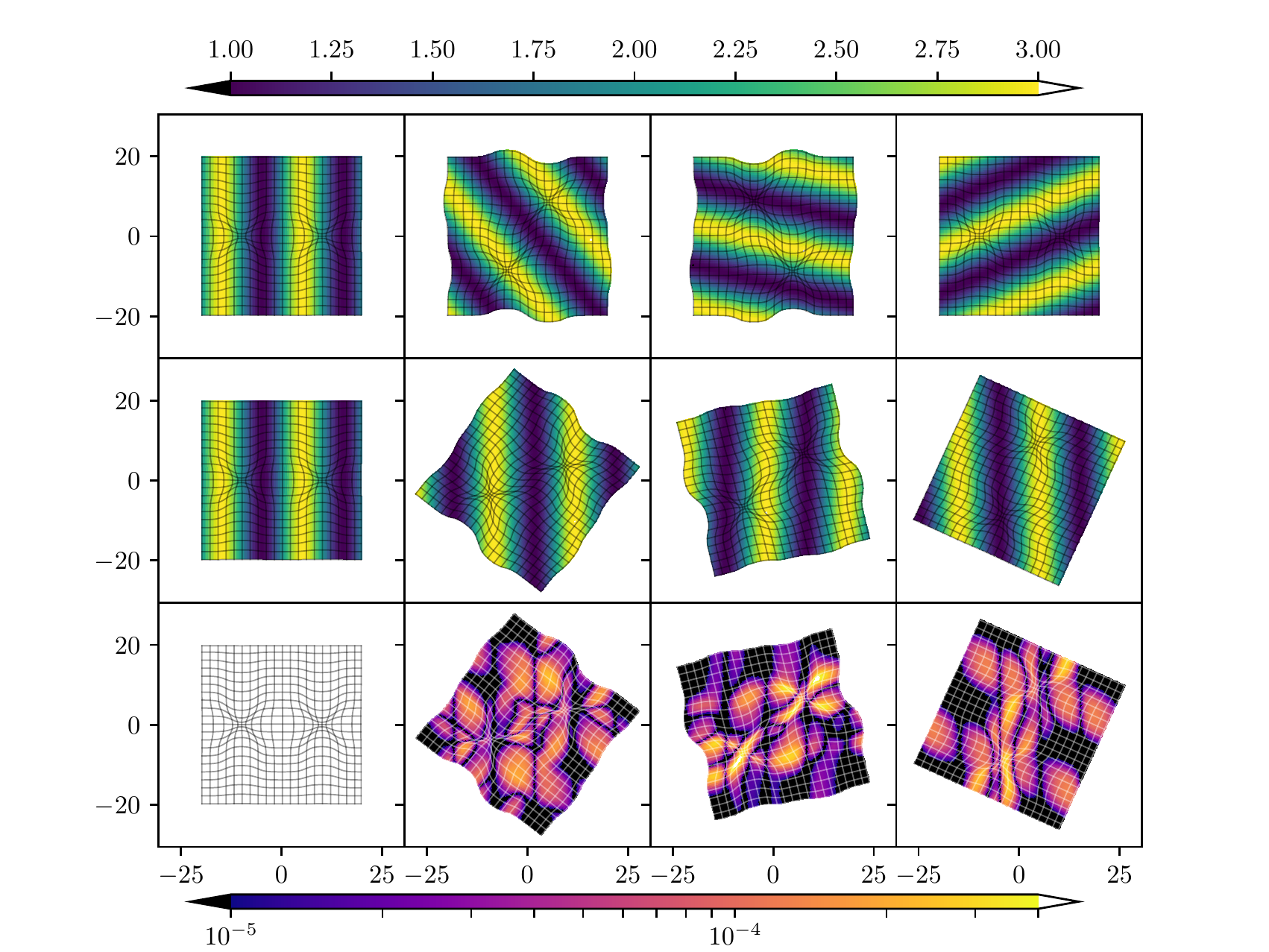}
  \caption{(Top Row) Time snapshots of the wave plotted with respect to the rotating frame coordinates used during the update.
  (Middle Row) Time snapshots of the wave plotted with respect to the fixed background coordinates.
  (Bottom Row) Logarithmic relative error in the wave plotted with respect to the fixed background coordinates. We do not plot
  the initial error because it is zero to machine precision.
  (Left to Right) Time snapshots correspond to $t = 0$, $t=66.5$, $t=133$, and $t=200$. We overlay
  every 5th gridline of the mesh.}
  \label{fig:wave_warp}
\end{figure}
We evolve for 5 wave crossing times of our simulation domain, or two complete rotations
of the warps, ($t = 200$) and plot snapshots 
of the evolution in Figure~\ref{fig:wave_warp}.
We find that even though the warps strongly distort the grid from the symmetry
of the wave solution, our relative error remains small (of order several $\times 10^{-4}$).
Furthermore, the top two rows, which show the evolution plotted against
the rotating frame coordinates and fixed background coordinates, highlight that the wave
propagation direction rotates counter-clockwise in the numerical coordinate basis where
the evolution is performed. This propagation direction time dependence is encoded
in the structure of $\Pi_{\mu}$. Finally, the values of $\Pi_\mu$ are further
altered and distorted to account for the warps in the mesh on which the wave
is updated.

\subsection{Test Two}
\label{sec:convergence}
While the original \pw code was designed to work with second
order time integration and linear interpolation for interpatch BCs
(recall the additional issue of interpolation of interpolated data
further reducing the convergence order),
we have modified \pw to be compatible with arbitrary order methods 
in space and time. 

We test the convergence rate of \pwWave using a Cartesian local patch
which is fully encompassed by a global Cartesian patch with a 
mesh refinement level of 2-to-1. The global patch cube extends from
$-20$ to $20$ in each dimension and the local patch cube extends
from $-10$ to $10$ in local coordinates $X_1^i$.
In all cases the resolution in each
dimension is equal and we employ a Courant-Friedrichs-Lewy (CFL) factor of 0.6 to determine the timestep.

We perform three tests; a
fixed local patch, a dynamically rotating local patch, and
a linearly translating local patch. For the fixed local patch,
we set the origin of the local patch to $\left(x,y,z\right) = \left(-1,-1,-1\right)$
to ensure that the local patch grid is offset from the global patch grid.
For the rotating local patch, we set the local patch origin at the global patch
origin and rotate counter-clockwise at a frequency of 0.01 about the z axis. For the translating
local patch, we set the initial local patch origin at $\left(x,y,z\right) = \left(-5,-5,0\right)$
and linearly translate with a velocity of $v^i = \left(0.1,0.1,0\right)$. In all cases we evolve
for 1 wave crossing time of the fixed local patch, $t=20$, which coincides with the wave period.

\begin{figure}[htb]
  \centering
  \includegraphics[width=0.65\textwidth]{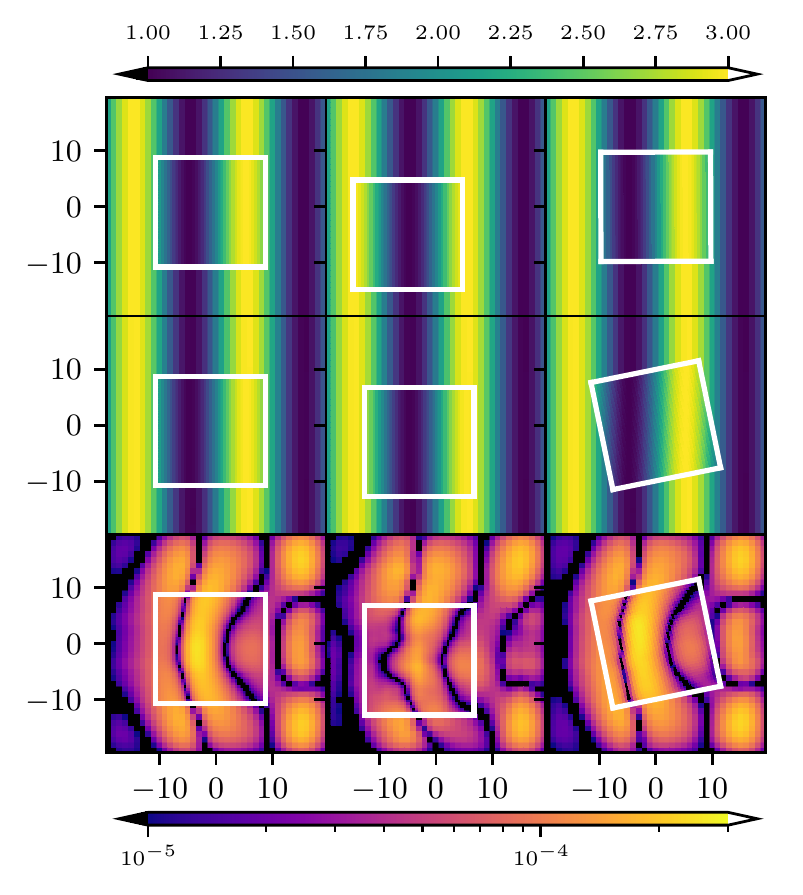}
  \caption{(Top Row) Initial configuration of the wave for our convergence
    study at the lowest resolution sampled.
    (Middle Row) Final configuration of the wave
    for our convergence study at the lowest resolution sampled.
    (Bottom Row) Logarithmic relative error in the wave at the end of
    the simulation for our lowest resolution sampled.
    (Left to Right) We plot the fixed local patch, translating local
    patch, and rotating local patch configurations. All plots
    are in the x-y plane at a slice of $z=0$. The edge of the local patch
    is denoted by the white box. Global patch resolution is $dx = dy = dz = 1$
    and local patch resolution is $dx = dy = dz = 0.5$.
    }
  \label{fig:convergence_snapshots}
\end{figure}
In Figure~\ref{fig:convergence_snapshots} we plot snapshots of the initial
and final configuration of the wave, as well as the relative error in the final state,
for each patch configuration at the lowest resolution sampled. Just like in the
warped configuration, 
we find that the relative error is small at the end of the simulation (of order several $\times 10^{-4}$)
even at a resolution of $dx = dy = dz = 1$ on global patch. Additionally, we note that the error
is continuous across the patch boundaries, showing negligible to no distortions due to
transportation onto and off of the mesh refinement of local patch. We note that the peak
of the wave has both transported off of and back onto the local patch during our test.
  
\begin{figure}[htb]
  \centering
  \includegraphics[width=0.65\textwidth]{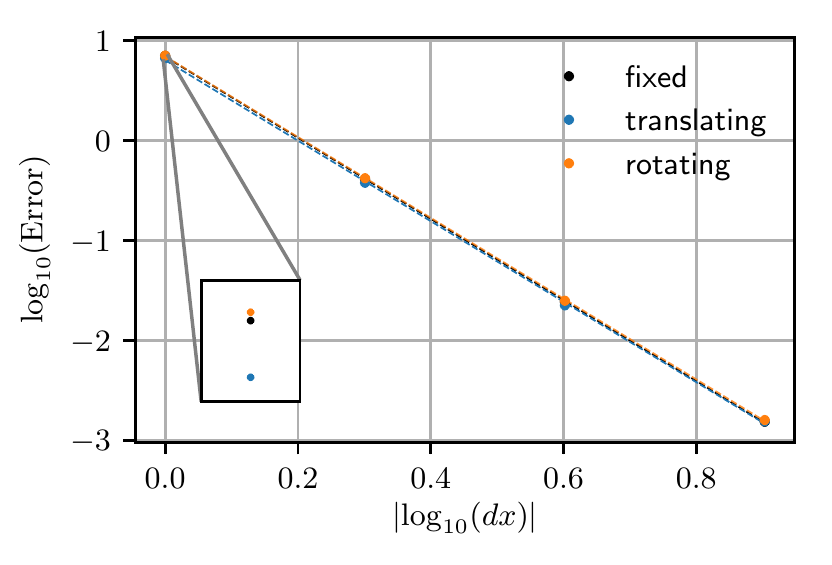}
  \caption{Total integrated relative error on both patches for our three convergence tests.
    Points denote the value for each resolution and the dashed line is a linear
    fit to the data. We include an inset of the lowest resolution points to highlight
    that they are distinctly separate. The size of the inset is not drawn to scale.
    The slopes are; fixed -4.06,  translating -4.03, and
    rotating -4.05.}
  \label{fig:convergence}
\end{figure}
To confirm the convergence rate of our algorithm, we sample 4 resolutions differing by a
factor of 2.
In such a configuration, for our globally 4th order scheme, the error should reduce by a factor
of 16 for each resolution increase.

We measure our convergence rate by calculating the total integrated error in the simulation domain
\begin{equation}
  \rm{Error} = \int \frac{\left| \phi - \phi_{\rm analytic} \right|}{\phi_{\rm analytic}} \sqrt{-g} \,d^3x,
\end{equation}
for all physically evolved cells on both patches. Here, $\sqrt{-g}$ is the determinant
of the metric tensor and generalizes the Jacobian term in front of the differentials.
In Figure~\ref{fig:convergence} we plot this integrated error
at the end of the simulation for all three tests on a log-log
scale. On this scale, the convergence order of the error is the linear slope of the data ($\times$ -1).

Two things are quickly apparent from the plot. First, the global convergence order of \pwWave
with our modifications is precisely the 4th order convergence rate we desired. Second, the global
integrated error between each simulation is comparable at all resolutions. This implies that the total error
of our scheme is relatively insensitive the location and trajectory of our refined local patch
for this test. Instead, the error is dominated by the 4th order spatial finite difference stencils
and time integration performed locally.
Furthermore, the effectiveness of our algorithm is exemplified by the rotating
patch mesh breaking with the symmetry of the plane wave in a time-dependent manner. Finally,
with the dynamic patch configurations we demonstrate the ability to dynamically
cover/uncover cells throughout the evolution with our modifications to interpatch ghost zone mapping
routines while preserving the desired convergence properties.

\subsection{Test Three}
\label{sec:dual_rotating}
We now perform a test which includes three patches
with one of each coordinate topology; Cartesian,
spherical, and cylindrical. In this test the local
patches both dynamically evolve in time with respect
to the global patch. Furthermore, we intentionally
configure the patches in a manner to make
interpatch reflections highly pronounced.

The global Cartesian grid has dimensions of 
$x \times y \times z = 80 \times 80 \times 40$
with its origin at $\left(x,y,z\right) = \left(0,0,0\right)$.
We sample the domain with $160 \times 160 \times 40$
cells. The spherical local patch is
centered at $\left(x,y,z\right) = \left(-22,0,0\right)$.
The uniform radial grid extends from $5$ to $20$, poloidal
angle from $\pi / 4$ to $3\pi / 4$, and we include
the full $2\pi$ in azimuthal angle. We sample the
spherical domain with $\left(r,\theta,\phi\right) = 30 \times 40 \times 200$
cells. The cylindrical local patch is centered
at $\left(x,y,z\right) = \left(22,0,0\right)$.
The radial domain and azimuthal domain extents
are the same as the spherical patch. The vertical
extent of the cylindrical patch is half that of
the global Cartesian patch. The cylindrical domain
is sampled with $\left(r,\phi,z\right) = 30 \times 200 \times 20$ cells.
Finally, the spherical patch rotates counter-clockwise
at a frequency of 0.01 and the cylindrical patch rotates
clockwise at a frequency of 0.03 (both with respect to their local z axes).

\clearpage
\begin{figure}[h!]
  \centering
  \includegraphics[width=.975\columnwidth]{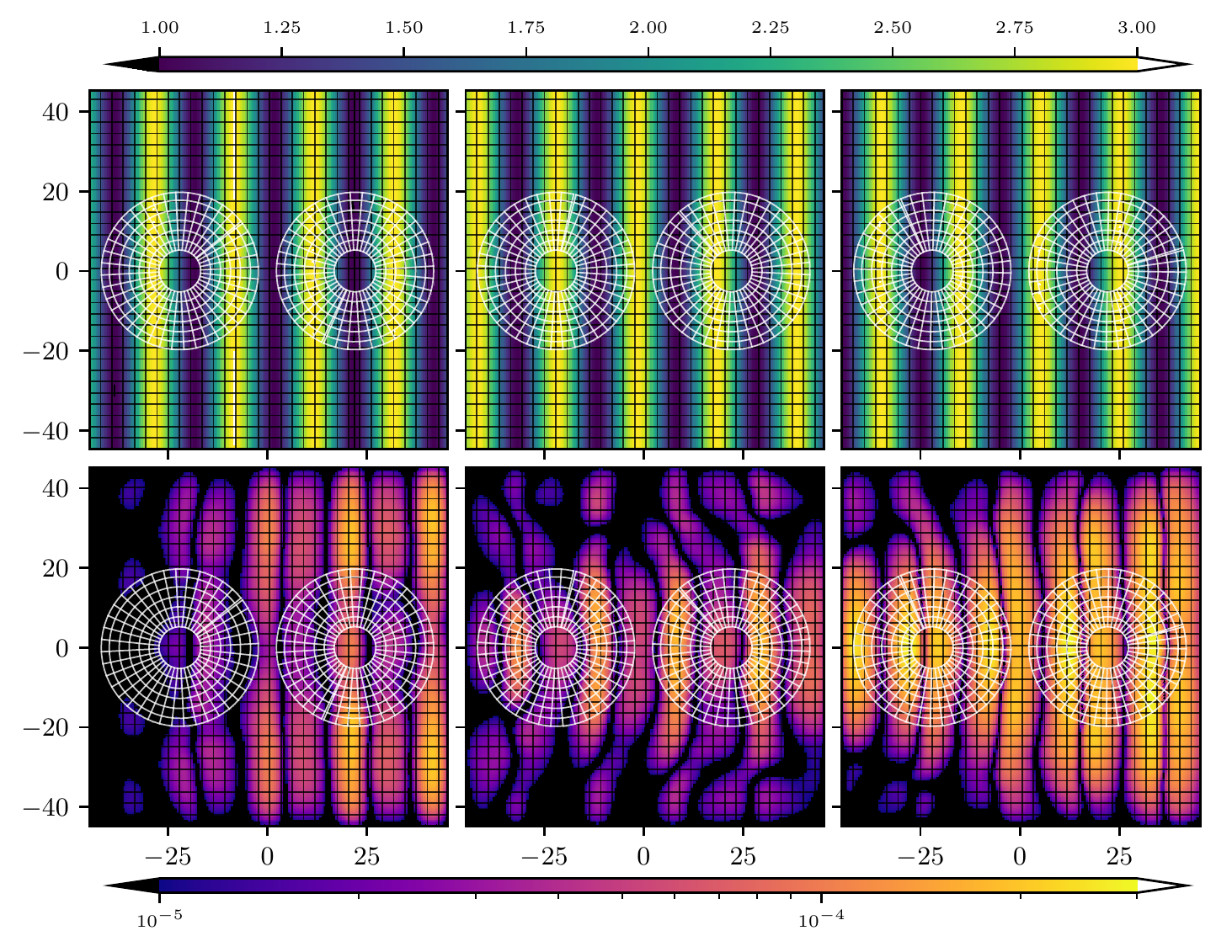}
  \caption{(Top) Snapshots in the x-y plane for the run at $z=0$. (Bottom) Relative error in the x-y plane at $z=0$.
    (Left to Right) Time snapshots correspond to $t=66.5$, $t=133$, and $t=200$. We overlay
    every 5th gridline of the meshes.}
  \label{fig:alltop_xy}
\end{figure}
\begin{figure}[h!]
  \centering
  \includegraphics[width=0.975\columnwidth]{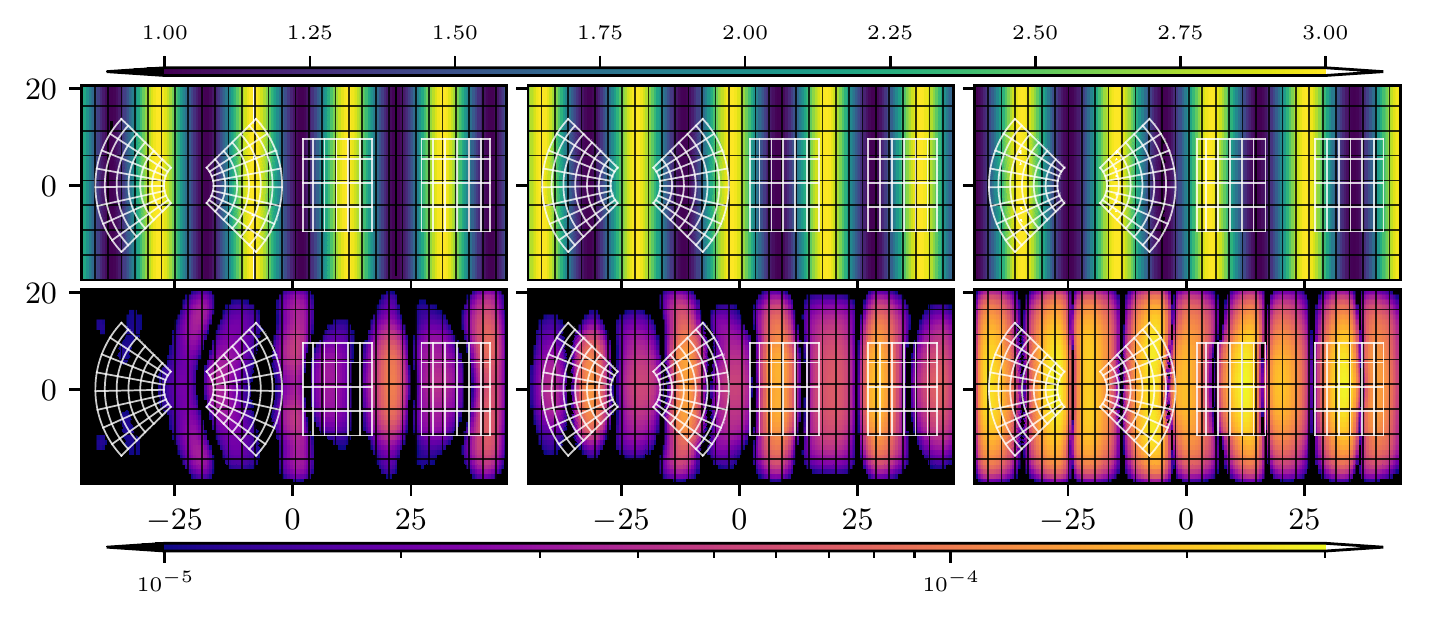}
  \caption{(Top) Snapshots in the x-z plane for the run at $y=0$. (Bottom) Relative error in the x-z plane at $y=0$.
    (Left to Right) Time snapshots correspond to $t=66.5$, $t=133$, and $t=200$. We overlay
    every 5th gridline of the meshes.}
  \label{fig:alltop_xz}
\end{figure}
\clearpage

In Figure~\ref{fig:alltop_xy} we plot snapshots of the wave and relative error
in the $x-y$ plane through our evolution. In these snapshots at $z=0$ the
resolution of the spherical and cylindrical meshes are identical. However,
the azimuthal phase of the local coordinates, along with
the Jacobians relating the local patch and global patch coordinates,
are different. Additionally, we plot the same snapshots in the $x-z$
plane in Figure~\ref{fig:alltop_xz}. The selection of $y=0$ for this
slice was made to represent one of the most error prone slices in the domain.
Observing the lower panels of Figures~\ref{fig:alltop_xy}~and~\ref{fig:alltop_xz} we find that the relative error
grows in time to $\mathcal{O}\left(10^{-4}\right)$. Had we continued the evolution, the error would likely
continue to increase.

However, this test was precisely designed to exacerbate errors associated with propagating the wave
through mesh refinement. For instance, the wave must cross mesh refinement boundaries
8 times for $\left|y\right| \lesssim 5$. Furthermore, no special care was taken to reduce the factor of cell
refinement between meshes in any portion of the domain and the local patches do not conform to
the symmetry of the plane wave. Finally, we intentionally placed the local patch edges immediately 
adjacent to the outer edges of the global patch along the direction of propagation. This last part is important because the analytic BCs
we apply act like a reflective boundary for any error modes generated by the mesh refinements and intrapatch truncation error.

Provided these considerations, we believe that the relative error growing to only $\mathcal{O}\left(10^{-4}\right)$
is a strong testament to the effectiveness of our interpatch BCs. We emphasize that any production simulation
employing the \pw infrastructure would likely have more transmissive boundary conditions and would place 
meshes to better capture the physics rather than infringe on the algorithm's ability to obtain
an accurate solution as we did here.

\subsection{Test Four}
\label{sec:alltop_moving}

To fully demonstrate \pwWave's mesh capabilities, we perform
another test which includes three patches with one of each coordinate topology.
In this test we include one translating patch, one rotating patch,
a curvi-linear global patch, and allow the local patches to overlap
in the buffer regions.

The global spherical grid extends from $r_{\rm in} = 5$ to $r_{\rm out} = 100$,
poloidally from $\pi/4$ to $3\pi/4$, and includes the full azimuthal domain.
We uniformly sample the domain with $r \times \theta \times \phi = 100 \times 100 \times 300$
cells. We include a Cartesian patch centered over the radial cutout which has dimensions
of $x \times y \times z = 30 \times 30 \times 20$ and sampled with $80 \times 80 \times 20$
cells. The patch rotates counter-clockwise in time at a frequency of 0.03 about the z-axis.
The cylindrical patch extends radially  from $r_{\rm in} = 5$ to $r_{\rm out} = 25$,
from -10 to 10 in the $z$ direction, and includes the full azimuthal angle.
When speaking of spherical coordinates we always refer to spherical radius. Conversely,
  when we are referring to a cylindrical coordinate system we refer to cylindrical radius.
We offset the initial position of the cylindrical patch from the global spherical
patch origin by $\left(x,y,z\right) = \left(45,45,0\right)$. We then translate
the cylindrical patch at a velocity of -0.5 in the x direction. We specify the 
simulation timestep to $dt = 0.05$.
\begin{figure}[htb]
  \includegraphics[width=\columnwidth]{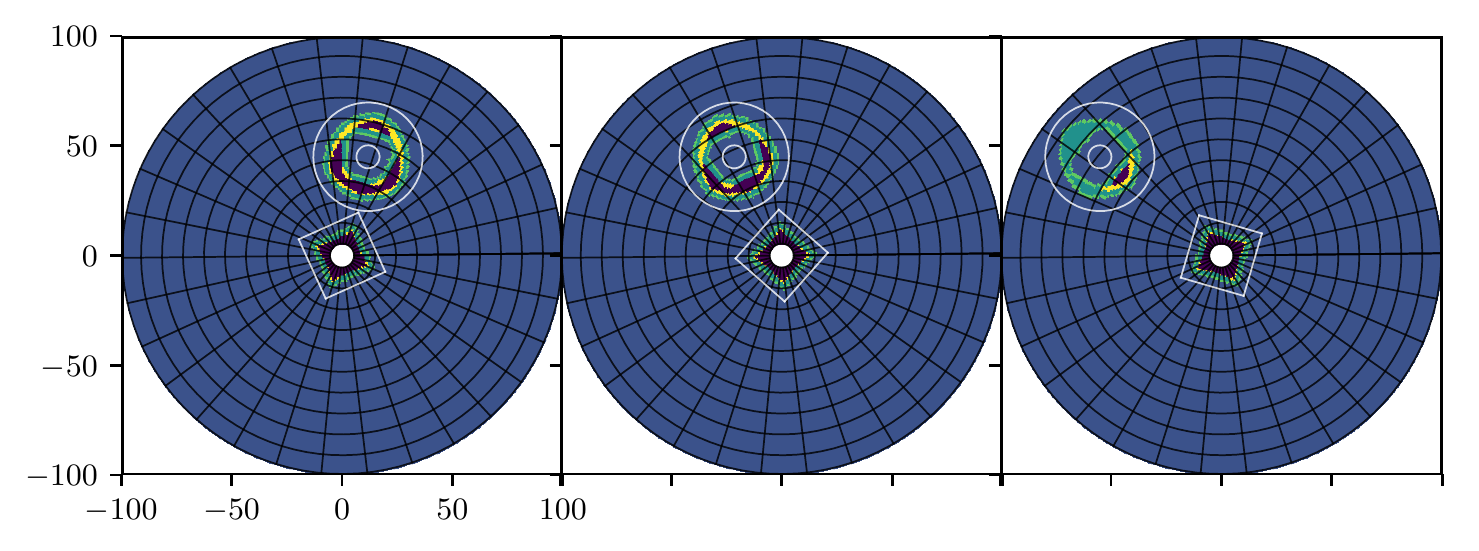}
  \caption{(Top) Snapshots of the interpatch interpolation flags in the x-y plane for the run at $z=0$.
    (Left to Right) Time snapshots correspond to $t=66.5$, $t=133$, and $t=200$. We overlay
    every 10th gridline of the global patch mesh and the outer physical boundaries of the local patch meshes.
    Purple denotes deactivated cells, blue evolved cells, and all other colors an interpatch interpolation zone.}
  \label{fig:alltop_moving_xy_cflag}
\end{figure}

We evolve two wave solutions in this setup until $t=200$, allowing the cylindrical patch to fully 
traverse the spherical global patch in the equatorial plane. Unlike the test 
presented in Section~\ref{sec:dual_rotating}, the interpolation flags identifying which 
zones require interpatch interpolated data must be dynamically updated in addition
to the coordinate mappings relating the patches. We demonstrate this in
Figure~\ref{fig:alltop_moving_xy_cflag}, where  we show that,
unlike Cartesian local patches which only have buffers cells outside the covered zones,
the cylindrical patch must have buffer cells extending from both of their inner radial boundaries.
Furthermore, the number of cells under the local cylindrical patch which may be deactivated
increases as the patch extends deeper into the spherical global patch radial domain.
This emphasizes that, in general, the amount of work a given patch must do, as well
as the structure of the deactivated, live, and interpolated cells, must be accounted
for dynamically and can vary significantly based on the mesh configurations at any timestep.

\subsubsection{Single Plane Wave}

\begin{figure}[htb]
  \includegraphics[width=\columnwidth]{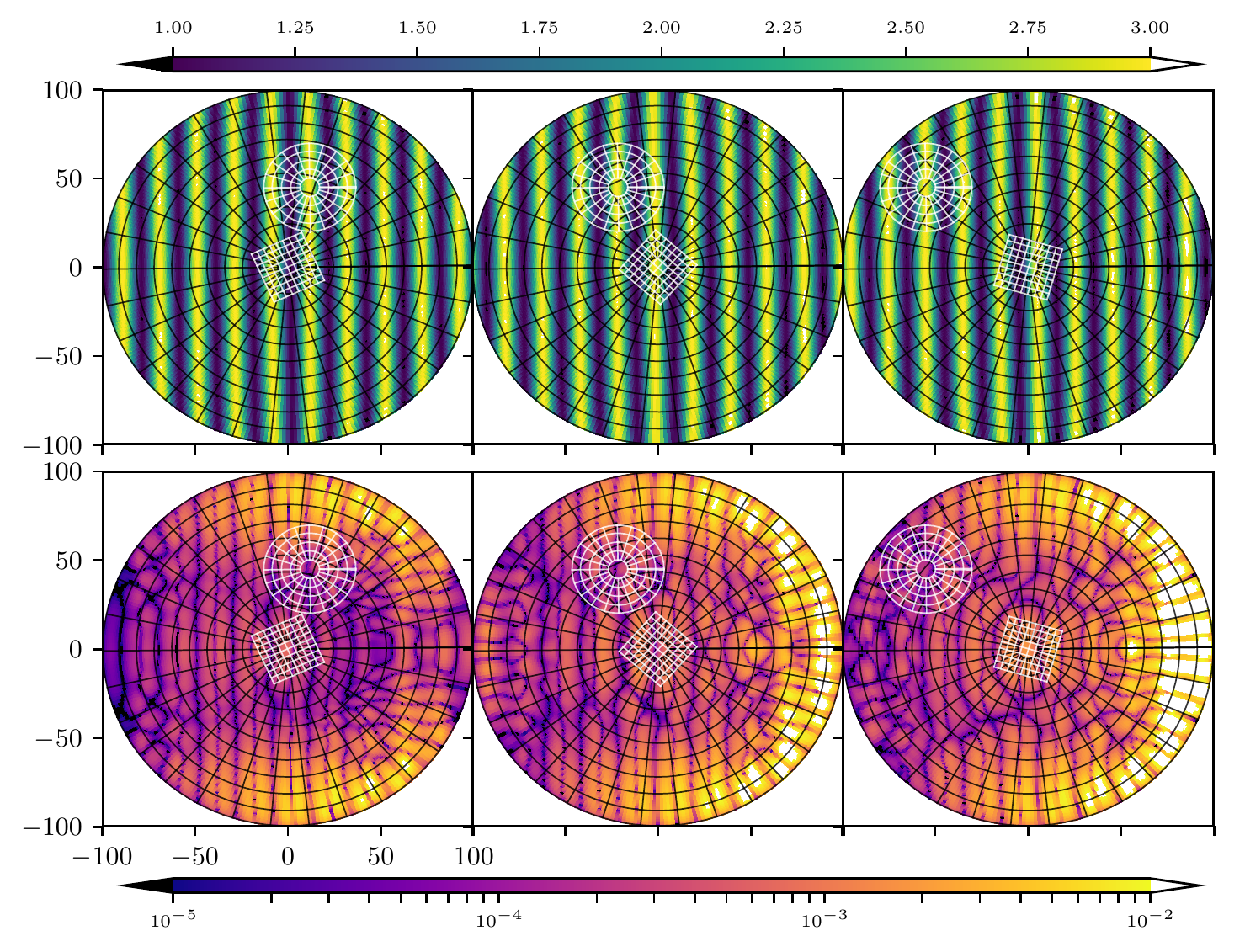}
  \caption{(Top) Snapshots in the x-y plane for the run at $z=0$. (Bottom) Relative error in the x-y plane at $z=0$.
    (Left to Right) Time snapshots correspond to $t=66.5$, $t=133$, and $t=200$. We overlay
    every 10th gridline of the meshes. }
  \label{fig:alltop_moving_xy}
\end{figure}

In Figure~\ref{fig:alltop_moving_xy} we plot snapshots of the simple 1D plane wave solution and relative error
in the equatorial $\theta = \pi/2$ plane of the global patch and $z=0$ plane of the local patches.
Unlike our previous simulations, we note that the maximal global error of the simulation
is significantly higher. However, we emphasize that this is due to the decreased effective
resolution of the global spherical patch at large radii. Furthermore, because error generated
in the very coarse cells at $\phi = \pi/2$ and $\phi = 3\pi/2$ ($dx \approx r d\phi \approx 2.1$)
cannot escape the domain with our analytic BCs, it is funneled along the outer radial boundary
to the $\phi \approx 0$ boundary where the plane wave propagates off the domain.

Furthermore, at $t \approx 75$, between snapshots 1 and 2 in the Figures, the corner of the
rotating Cartesian local patch penetrates through the cylindrical outer boundary. At these times
the physical ghost zones from both patches draw interpolated data from the buffer regions under
the other local patch. While 
our algorithm does not allow live or buffer cells in local patches to be covered by other local patches, 
it does allow their ghost zone regions to overlap.

\subsubsection{Superposition of Plane Waves}
\begin{figure}
  \includegraphics[width=\columnwidth]{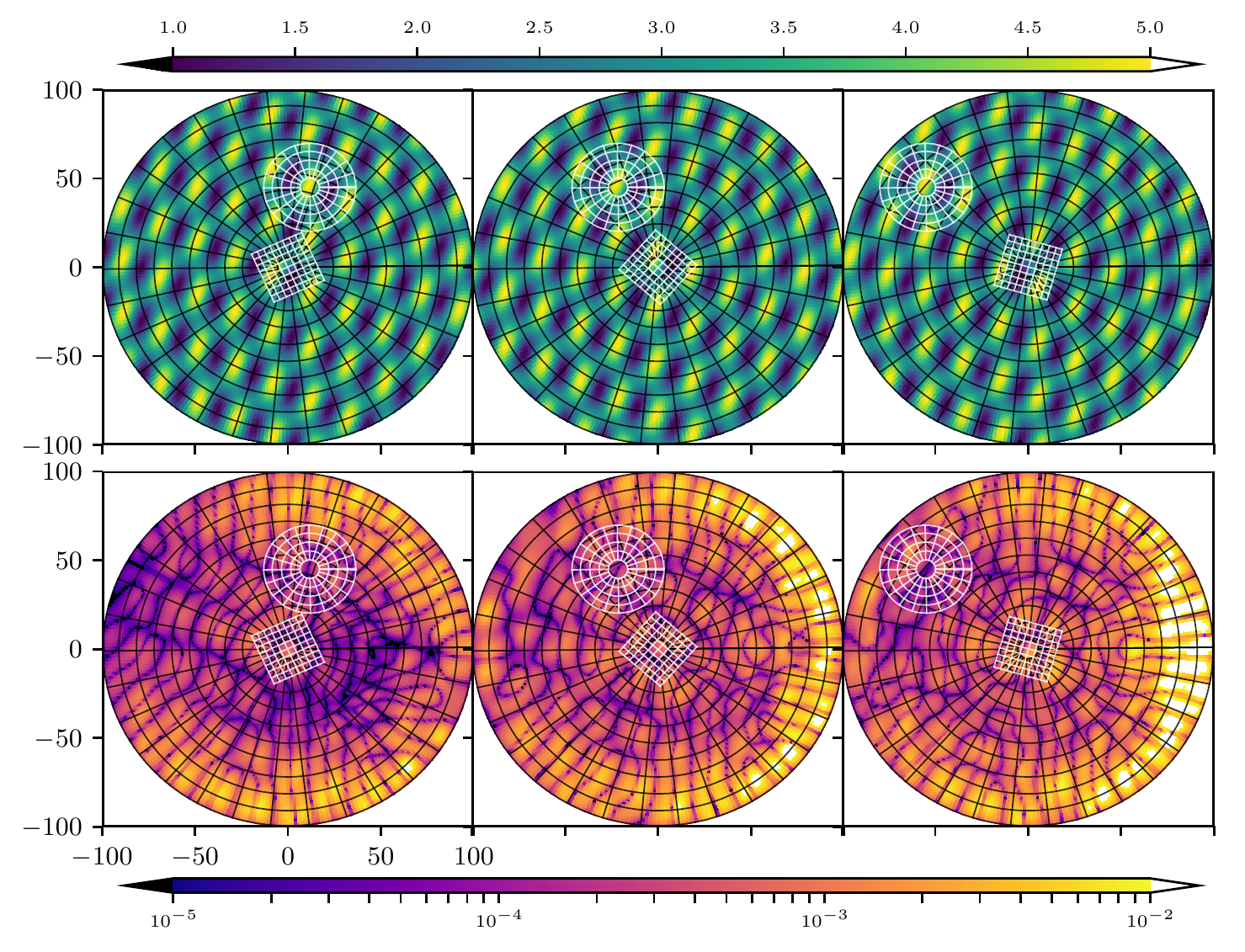}
  \caption{(Top) Snapshots in the x-y plane for the run at $z=0$. (Bottom) Relative error in the x-y plane at $z=0$.
    (Left to Right) Time snapshots correspond to $t=66.5$, $t=133$, and $t=200$. We overlay
    every 10th gridline of the meshes. }
  \label{fig:alltop_moving_xy_superpose}
\end{figure}

Finally, to demonstrate that the 1D nature of our simple plane wave
is not exploited by our code, we evolve a superposition of two plane
waves in the same patch configuration. Each wave has a different
wave vector and frequency. Specifically, we evolve
\begin{eqnarray}
  \phi\left(t,x,y,z\right) &=& \sin\left[\frac{2\pi}{L_1}\left(x-t\right)\right] 
   + \cos\left[\frac{2\pi}{L_2}\left(x - y + z - \sqrt{3}t\right)\right] + 3,
\end{eqnarray}
where we set $L_1 = 20$ and $L_2 = 30$. This solution requires resolving
full 3D wave propagation and two independent wave periods simultaneously.
We plot a slice at $z=0$ for the same time snapshots as in the previous section
in Figure~\ref{fig:alltop_moving_xy_superpose}.

In the figure we note that the wave features which our code must resolve are quite
different from the single plane wave solution. However, by comparing
snapshots in
Figures~\ref{fig:alltop_moving_xy}~and~\ref{fig:alltop_moving_xy_superpose},
we observe that the overall magnitude of the error is comparable. This
demonstrates that, so long as the grid is sufficiently sampled to resolve
the gradients of the solution, our error is dominated by the truncation
error of our spatial and temporal discretization. 

\section{Conclusions}
\label{sec:conclusions}
We have described extensions to the \pw ~\citep{PATCHWORK} and enhanced \pw ~\citep{PWMHD} (upon which this code is built) framework necessary for 
this study, high-order finite difference methods, as well as those developed jointly for this publication and with the new 
MHD-capable code \pwmhd \citep{PWMHD}. 
We have additionally 
described in detail some of the underlying enhanced \pw infrastructure for clarity. 

We have extended the framework
to be \textit{multimethod} compatible, improved the accuracy of interpatch
BCs, and removed assumptions pertaining to finite volume fluid evolutions.
These extensions were motivated by a multimethod astrophysics application
coupling numerical solutions to the EFE via finite difference methods,
to the evolution of the MHD equations employing finite volume methods. As 
the original \pw has already been successfully employed for 
fluid evolutions in ~\cite{PATCHWORK}, and its MHD counterpart is currently 
being deployed ~\citep{PWMHDL}, we have demonstrated only the finite difference
component here through the evolution of a scalar wave toy-model. We implemented
this toy-model into the \harm code.

While there is certainly MPI overhead associated with our
infrastructure, the exact cost is difficult to predict, as well as
often specific to the simulation or application. For instance,
the relative expense of interpatch interpolation to the time integration
of the state vector, the number of cells which require interpolation,
the fraction of deactivated cells on a patch, the cell count disparity between various
patches, and each patch's domain decomposition  all play a role. 
We therefore do not expect
the overhead associated with our toy-model to necessarily be a good
predictor for another application.
As such, we have left a full discussion
of the scaling and overhead to a future publication which
is optimized to its target application.  

However, we emphasize
that the extreme flexibility of \pw allows users to develop meshes that
are custom tailored to minimize cell count, maximize timestep, and
completely remove many-level nested box-in-box AMR. These
benefits, when combined with arbitrary patch motion,
have the potential to far outweigh the overhead costs. For instance,
\pwmhd is currently being deployed to efficiently extended the simulation of SMBBH accretion
presented in ~\cite{Bowen18,Bowen19} from 12 to 30 binary orbital
periods at a fraction of the computational cost ~\citep{PWMHDL}.
Furthermore, because each application code is itself parallelized, users may 
domain decompose their application code as they see fit to
limit communication overhead.

In addition to the simplicity of implementation, a second order formulation of
the scalar wave has stability criteria similar to many other 
applications in computational physics. Applications
of wave-like systems span general relativity to quantum mechanics. 
Although we elected to implement a coordinate invariant form of the wave equation
using a metric tensor, we emphasize that this is \textit{not} a critical
component of the new \pw infrastructure. Furthermore, our code would be
equally capable of evolving another set of hyperbolic PDEs
at arbitrary order. The only necessary ingredients are
a time integrator for the equations, an interpolator, a method of identifying
neighboring cells, and coordinate
transformation rules for the state vector. 

Our expansions
allow for the new enhanced \pw code (and hence both \pwWave and \pwmhd)  
to be compatible with a wide range of
infrastructures beyond finite difference or finite volume.
For instance, our infrastructure could be coupled to any number of 
integration methods, where each application provides the necessary
methods listed above.
In principle, the code we present here could even be coupled to
non-hyperbolic PDEs, such as an elliptic multigrid solver, where
\pw would provide intergrid interpolation at some iteration
frequency.

Through our toy-model, we have demonstrated that the further enhanced \pw infrastructure may be 
deployed to produce a generic multipatch scheme which
is globally convergent at the order of the mesh/temporal discretization.
This is facilitated by our improvements to the interpatch interpolation,
introduction of an arbitrary size buffer region, and the capability of handling
multiple state vectors simultaneously (each employing their own interpatch
interpolation/transformation procedures).
With these extensions, \pw promises 
to provide accurate solutions for physics applications which have disparate
resolution/physics requirements, different numerical technique requirements,
and allows for moving mesh advantages even in the presence
of complex  solution geometries.
Finally, this proof of principle calculation represents
a significant step towards the goal of time-dependent multipatch schemes
in numerical relativity.


\section*{Acknowledgments}
D.~B.~B. is supported by the US Department of Energy through the Los Alamos 
National Laboratory. Los Alamos National Laboratory is operated by Triad National 
Security, LLC, for the National Nuclear Security Administration of U.S. 
Department of Energy (Contract No. 89233218CNA000001).
For this work, D.~B.~B. also acknowledges support from NSF grants
AST-1028087, AST-1516150, PHY-1707946, OAC-1550436 and
OAC-1516125. 
M. A. is a Fellow of the RIT's Frontier of Gravitational Wave Astronomy.
M.~A., M.~C., V.~M. and Y.~Z. acknowledge support from AST-1028087, 
AST-1516150, PHY-1912632, PHY-1707946, PHY-1550436, OAC-1550436 and
OAC-1516125, and from NASA TCAN grant No. 80NSSC18K1488.
V.~M. was partially supported by the Exascale Computing Project (17-SC-20-SC), 
a collaborative effort of the U.S. Department of Energy Office of 
Science and the National Nuclear Security Administration.
S.~C.~N. was supported by AST-1028087, AST-1515982 and
OAC-1515969, and by an appointment to the NASA Postdoctoral Program at
the Goddard Space Flight Center administrated by USRA through a
contract with NASA.
The work of R.~M.~C. was funded under the auspices of Los Alamos National Laboratory, 
operated by Triad National Security, LLC, for the National Nuclear Security 
Administration of U.S. Department of Energy (Contract  No. 89233218CNA000001).
J.~H.~K. was partially supported by NSF Grants PHYS-1707826 and AST-1715032.

Computational resources were provided by the Blue Waters
sustained-petascale computing NSF projects OAC-1811228 and OAC-1516125. 
Blue Waters is a joint effort of the University of Illinois at Urbana-Champaign and its
National Center for Supercomputing Applications. Additional resources
were provided by the RIT's BlueSky and Green Pairie Clusters   
acquired with NSF grants AST-1028087, PHY-0722703, PHY-1229173 and PHY-1726215.

\appendix
\section{Coordinate Invariant Wave Equation}
\label{appendix:tensor_calc}
In order to understand how our code handles the generalization of motions
and coordinates, a basic understanding of tensor calculus on manifolds is required.
We present here a very basic introduction for readers unfamiliar with
tensor calculus so as to understand the origins of the equations presented in
Section~\ref{sec:eom}. For more background information see ~\cite{1973grav.book.....M}.

Consider a 4-dimensional Riemannian manifold (time + space). This manifold may have
arbitrary curvature and may be spanned using any freely specifiable coordinate
basis vectors $e_\mu$. In this situation, the differential distance
between points $(ds^2)$ may be expressed in a coordinate invariant manner as
\begin{equation}
  ds^2 = \sum_{\mu = 1}^{4}\sum_{ \nu = 1}^{4}g_{\mu \nu} dx^{\mu} dx^{\nu},
\end{equation}
where repeated upper and lower Greek indices always imply a sum over all 4 spacetime indices
(from here we drop the explicit summation symbol) and $g_{\mu \nu}$ is the metric tensor of the
4-dimensional manifold. In the Cartesian representation of flat (Minkowski)
spacetime, our indexing corresponds to $x^{\mu} = \left(t,x,y,z\right)$ and the
metric tensor, $g_{\mu \nu} = \textrm{diag}\left(-1,1,1,1\right)$, recovers the standard
Lorentzian displacement  formula: $\left | \vec{x}^\prime - \vec{x} \right| = \sqrt{ -\left(x_1^\prime - x_1\right)^2 +  \sum_{i=2}^4 \left(x_i^\prime - x_i\right)^2 }$. 
The metric tensor obeys the standard tensorial transformation law of
\begin{equation}
  g_{\mu^\prime \nu^\prime} = \frac{\partial x^\mu}{\partial x^{\mu^\prime}}\frac{\partial x^\nu}{\partial x^{\nu^\prime}} g_{\mu \nu}.
\end{equation}
It may be trivially shown that for spherical $\left(t,r,\theta,\phi\right)$ and cylindrical
$\left(t,r,\phi,z\right)$ coordinates the flat spacetime (Minkowski) metric tensor becomes
$g_{\mu \nu} = \textrm{diag}\left(-1,1,r^2,r^2\sin\theta\right)$ and
$g_{\mu \nu} = \textrm{diag}\left(-1,1,r,1\right)$ respectively.

Additionally, in deriving our EOM we generalize the partial derivative operator.
The generalized form of differentiation, the covariant derivative, correctly accounts
for curvilinear scale factors as well as curvature. By definition, the covariant
derivative acting on the metric tensor is always zero. Under this assumption,
the covariant derivative acting on a scalar function is precisely the partial
derivative operator. However, for tensorial quantities (such as $\partial_\mu \phi = \Pi_\mu$)
it may be shown that the derivative of an arbitrary tensor of rank $n \times m$ along coordinate $x^\mu$ is
\begin{eqnarray}
  \nabla_{\mu} {T^{\alpha_1 \cdots \alpha_n}}_{\beta_1 \cdots \beta_m} &=& \partial_{\mu} {T^{\alpha_1 \cdots \alpha_n}}_{\beta_1 \cdots \beta_m} \\
& &\, + {\Gamma^{\alpha_1}}_{ \mu \lambda} {T^{\lambda \alpha_2 \cdots \alpha_n}}_{\beta_1 \cdots \beta_m} + {\Gamma^{\alpha_2}}_{ \mu \lambda} {T^{\alpha_1 \lambda \cdots \alpha_n}}_{\beta_1 \cdots \beta_m} + \cdots + {\Gamma^{\alpha_n}}_{ \mu \lambda} {T^{\alpha_1 \alpha_2 \cdots \lambda}}_{\beta_1 \cdots \beta_m}\nonumber\\
& &\, - {\Gamma^{\lambda}}_{ \mu \beta_1} {T^{\alpha_1 \cdots \alpha_n}}_{\lambda \beta_2 \cdots \beta_m} - {\Gamma^{\lambda}}_{ \mu \beta_2} {T^{\alpha_1 \cdots \alpha_n}}_{\beta_1 \lambda \cdots \beta_m} - \cdots - {\Gamma^{\lambda}}_{ \mu \beta_m} {T^{\alpha_1 \cdots \alpha_n}}_{\beta_1 \beta_2 \cdots \lambda}.\nonumber
\end{eqnarray}
Here ${\Gamma^\alpha}_{\mu \nu}$ are the Christoffel symbols of the second kind, or connection coefficients,
and may be expressed in terms of the metric tensor as
\begin{equation}
  {\Gamma^\alpha}_{\mu \nu} = \frac{1}{2}g^{\alpha \lambda}\left( \partial_\mu g_{\lambda \nu} + \partial_\nu g_{\mu \lambda} - \partial_\lambda g_{\mu \nu} \right).
\end{equation}
$g^{\mu \nu}$ is the inverse metric tensor which may be obtained by taking the inverse
matrix representation of the $4\times4$ metric tensor $g_{\mu \nu}$. 

In deriving and relating our expressions, it is important to note that the location
of indices (upper vs lower indices) in this notation must be respected. The traditional vector representation
of a one-form $\Pi_{\mu}$ may be found by ``raising'' the index with the
metric tensor
\begin{equation}
  \Pi^{\mu} = g^{\mu \lambda} \Pi_{\lambda}.
\end{equation}
Conversely, the one-form may be found from the vector by ``lowering'' the index with the
metric tensor
\begin{equation}
  \Pi_{\mu} = g_{\mu \lambda} \Pi^{\lambda}.
\end{equation}
This may be performed for any number of indices, with each index obeying one of the
two expressions above.

Finally, covariant expressions may be obtained by replacing partial derivatives with
covariant derivatives. That is, the laplacian operator in Cartesian
coordinates $\nabla^2 = \sum_i \partial_i^2 = \partial^i \partial_i$ may be obtained
by first lowering the first index 
\begin{equation}
  \nabla^2 = g^{ij}\partial_i \partial_j,
\end{equation}
where repeated latin indices only sum over spatial indices 
of the manifold ($(x,y,z)$, $(r,\theta,\phi)$, $(r,\phi,z)$ etc.). Next, 
partial derivatives are replaced with their covariant generalization producing
\begin{equation}
  \nabla^2 = g^{ij}\nabla_i \nabla_j.
\end{equation}
Finally, for convenience,
we set the wave speed to the speed of light. We may then absorb the
$-\partial_t \partial_t \phi$ term into a 4 dimensional sum with the metric tensor
and define the fully coviariant wave expression presented in Equation~\ref{eq:covariant_wave}.

\hfill \break
\bibliography{bhm_references}

\end{document}